\documentclass{article}
\usepackage{amssymb,amsbsy,amsmath,amsfonts,amssymb,amscd}
\usepackage{graphicx,epsfig,times,epsf,subfigure, cases}%,harvard}

\newcommand{\N}{\mathbb{N}}
\newcommand{\Z}{\mathbb{Z}}

\newcommand{\R}{\mathbb{R}}
\newcommand{\C}{\mathbb{C}}

\newcommand{\brB}{\mathbf{B}}
\newcommand{\brD}{\mathbf{D}}
\newcommand{\brE}{\mathbf{E}}

\newcommand{\brH}{\mathbf{H}}
\newcommand{\brJ}{\mathbf{J}}
\newcommand{\brM}{\mathbf{M}}
\newcommand{\brP}{\mathbf{P}}

\newcommand{\bre}{\mathbf{e}}

\newcommand{\brs}{\mathbf{s}}
\newcommand{\brv}{\mathbf{v}}

\newcommand{\eps}{\varepsilon}

\newcommand{\RE}{\Re\it{e}}
\newcommand{\IM}{\Im\it{m}}

\newcommand{\dfl}{\lfloor}
\newcommand{\ffl}{\rfloor}

\begin{document}

\title{Optical scattering by a nonlinear medium, I: from Maxwell's equations to numerically tractable equations}

\author{Pierre Godard, Fr\'ed\'eric Zolla, Andr\'e Nicolet \vspace{.2cm} \\
\small{Institut Fresnel UMR CNRS 6133, Facult\'e de Saint-J\'er\^ome case 162} \\
\small{13397 Marseille Cedex 20, France} \\
\small{pierre.godard@fresnel.fr,frederic.zolla@fresnel.fr,andre.nicolet@fresnel.fr}}

\date{}

%\affiliation{Institut Fresnel UMR CNRS 6133, Facult\'e de Saint-J\'er\^ome case 162 \\
%13397 Marseille Cedex 20, France \\
%pierre.godard@fresnel.fr,frederic.zolla@fresnel.fr,andre.nicolet@fresnel.fr}

\maketitle

\begin{abstract}
A new method to find the propagation equation system governing the
scattering of an electromagnetic wave by a nonlinear medium is
proposed. The aim is to let the effects appear spontaneously,
deleting as far as possible the phenomenological ideas. In this way,
we obtain propagation equation systems that encodes several
nonlinear effects. Once these systems obtained, the numerical values
of the tensors that characterize the answer of the medium to an
electromagnetic perturbation give weights to the different
effects.\newline%
This aim is partly reached in this study, especially when treating
harmonic generations. For this, we start from the Maxwell's equation
and give rigorously all the hypothesis needed to attain equation
systems that can be solved, at least from a numerical point of view.
Finally, a symmetry on the susceptibility tensors that ensures that
a medium is lossless from the electromagnetic point of view is
worked out.
\end{abstract}

\section{Introduction}

The usual route to treat nonlinear media from the electromagnetic
point of view is to expand the electric permittivity as a function
of the electric field. These expansions quickly get huge and so one
has first to select the effect we want to consider, and then
systematically simplify the equations, keeping only the terms that
contribute to this effect (\cite{bBloembergen65,Boyd}). A look at
the literature shows that the precision that can be obtained with
this method is out of question.

Nevertheless, it would appear more satisfactory to start only with
Maxwell's equations and to let the effects appear by themselves,
introducing a systematical way to simplify the equations. The
simplification is thus \emph{a priori}, generic ; the relative
importance of the observed effects being only \emph{a posteriori}
determined, by the numerical values of the susceptibility tensors.
This paper aims at drawing this new road.

For this, some general work on the constitutive relations had to be
done. We expose it in the subsection \ref{SubsecIndFields}. Then,
with some assumptions, we recover in the next subsection the
standard expression of the electric susceptibility tensors. Most of
these assumptions are justified for the presentation of this paper,
and avoiding them would not have lead to insurmountable
difficulties.

Having expressed the answer of a medium, we obtain the propagation
equation system satisfied by the electric field in
\ref{SubsecGenProgEqSyst}. This system, parameterized by a
continuous parameter, is far too complicated to be solved, even with
numerical methods. This leads us to partly leave our general aim,
and concentrate on harmonic generation. In this context, we
introduce a new notion, called the degree: it determines, in generic
terms, which interaction between the Fourier components of the
electric field have to be taken into account. The subsections
\ref{SubsecSecOrder} and \ref{SubsecThirdOrder} are devoted to the
propagation equation systems in the lowest order of nonlinearity and
in the lowest degree; the physical effects are then clearly
identified. We hope the obtention of these equation systems from
\emph{ab initio} principles will convince the reader about the
interest of the method.

In a second part (section \ref{SectionEnergy}), we try to answer, we
think more rigorously that what can be found in the literature, to
the question: \`\,how the electric energy variation is determined by
the susceptibility tensor fields of a medium?' If the general answer
is still unknown, we give a sufficient condition for a medium to be
lossless.

%The aim of this paper is twofold: first, we want to deduce as
%clearly as possible the effects encountered in nonlinear optics from
%Maxwell's equations; and we want to introduce a systematical way for
%simplifying the equations. For the time being, we have to admit that
%this method is not completely over; but we hope that, when this
%\emph{systematic} way will be fully applied, we could be sure that
%all the nonlinear effects, as the Kerr-optical effect, the harmonic
%or subharmonic generations, the Raman or Brilloin scattering, etc.
%will be known.

\vspace{.2cm}

This paper is intended to be appreciated by the theoretical
physicists, either working in the nonlinear optics field or not, as
well as the applied physicists. To this aim, we start from general
concepts and end with equation systems that can be numerically
solved. Having clarified the way these systems appear, we have
realized some simulations, one of them appearing in the companion
paper \cite{aGodard11b}. In particular, these simulations confirmed
our prediction on the energy criterion.

\section{The electromagnetic constitutive relations}

\subsection{A tractable expression for the inductive
fields}\label{SubsecIndFields}

As said in the introduction, the aim of this article is to expose a
new route to the equations of nonlinear optics, starting only with
the Maxwell's equations. We will work at a completely classical
level, even if, of course, it is hoped that the effective
characteristics of the medium will in future works clearly come from
a microscopic point of view. In other words, in this study, the
characterization of a medium is considered to be fully determined as
soon as a set of (susceptibility) tensors are given.

Let us start with the Maxwell's equations. We consider here that the
primitive fields are the electric field $\brE$ and the magnetic
field $\brB$. The first set of equations gives a `coherence'
condition between $\brE$ and $\brB$:

\begin{equation}\label{EqMaxPrim}
\left( \begin{array}{c}
  \nabla\times\,\brE+\partial_t\brB \\
  \nabla\cdot\,\brB
\end{array} \right)=\left( \begin{array}{c}
  0 \\
  0
\end{array} \right).
\end{equation}
The second set of equations links the induction fields with the
sources of the electromagnetic fields. We thus denote by $\brD$ the
electric induction, $\brB$ the magnetic induction, $\rho$ the charge
density and $\brJ$ the current density. Then, the second set of
Maxwell's equations is

\begin{equation}\label{EqMaxIndSource}
\left( \begin{array}{c}
  \nabla\times\,\brH-\partial_t\brD \\
  \nabla\cdot\,\brD
\end{array} \right)=
\left( \begin{array}{c}
  \brJ \\
  \rho
\end{array} \right).
\end{equation}

Supposing that the sources are known, we are looking for the
electromagnetic fields. It appears that we have to add some
hypothesis to answer this problem, for, in each point $\brs$ of
$\R^3$ and for any time $t$, we have twelve unknowns (the three
components of the four vectors $\brE(\brs,t)$, $\brB(\brs,t)$,
$\brD(\brs,t)$ and $\brH(\brs,t)$), but the Maxwell's equations give
only two vectorial and two scalar equalities. The set of Maxwell's
equations being independent of the medium in which the fields
oscillate, we have to encode the (electromagnetic) characteristics
of the medium, with two (vectorial) relations that give the
inductive fields in function of the primitive fields. We pose the
following set of equations:

\begin{numcases}{\,}
        \brD(\brs,t)=\mathfrak{D}\big(\{\brE(\mbox{\boldmath$\sigma$},\tau)\}_{(\mbox{\boldmath$\sigma$},\tau)\in S_e^d(\brs)\times T_e^d(t)},\{\brB(\mbox{\boldmath$\sigma$},\tau)\}_{(\mbox{\boldmath$\sigma$},\tau)\in S_b^d(\brs)\times T_b^d(t)},\brs,t\big) \label{EqConstitutionGeneralesD} \\
        \brH(\brs,t)=\mathfrak{H}\big(\{\brE(\mbox{\boldmath$\sigma$},\tau)\}_{(\mbox{\boldmath$\sigma$},\tau)\in
        S_e^h(\brs)\times T_e^h(t)},\{\brB(\mbox{\boldmath$\sigma$},\tau)\}_{(\mbox{\boldmath$\sigma$},\tau)\in S_b^h(\brs)\times
T_b^h(t)},\brs,t\big)\label{EqConstitutionGeneralesH}
\end{numcases}
where, for example,
$\{\brE(\mbox{\boldmath$\sigma$},\tau)\}_{(\mbox{\boldmath$\sigma$},\tau)\in
S_e^d(\brs)\times T_e^d(t)}$ is the set of all values of the
electric field at the place \mbox{\boldmath$\sigma$} and time
$\tau$, when \mbox{\boldmath$\sigma$} and $\tau$ run over the sets
$S_e^d(\brs)\subset\R^3$ and $T_e^d(t)\subset\R$ respectively. As
said above, the `fabrication' of the electromagnetic induction is
done at another scale than the one considered here, and we are
reduced to see it like a black box: at the input are, for each place
$\brs$ and each time $t$, the primitive fields, and $\brs$ and $t$;
at the output are the inductive vectors $\brD(\brs,t)$ and
$\brH(\brs,t)$. Several authors (\cite{aKarlsson92,bWeiglhofer03},
see also \cite{Nicolet}) give expressions similar to
(\ref{EqConstitutionGeneralesD}-\ref{EqConstitutionGeneralesH}), but
we think that the one we present is more rigorous and allows a clear
transcription of the properties of the medium. The article
\cite{aMontanaro09}, and some references found here, is not
restricted to pure electromagnetic phenomena.

The transcriptions, between the properties of the medium and the
restriction it induces on the functionals $\mathfrak{D}$ and
$\mathfrak{H}$ are the following one:

\begin{itemize}
\item The causality of the medium implies that

$$T_a^b(t)\subset(-\infty,t],\quad(a,b)\in\{e,b\}\times\{d,h\},$$
which means that the times $\tau$ at which are evaluated the
electric and magnetic fields that contribute to $\brD(\brs,t)$ and
$\brH(\brs,t)$ are prior to $t$.

\item The locality (in space) of the medium implies that

$$S_a^b(\brs)=\{\brs\},\quad(a,b)\in\{e,b\}\times\{d,h\}.$$
Allowing the notations' abuse that consists in keeping the same
names for the inductive functionals, the new expression are

\begin{numcases}{\,}
        \brD(\brs,t)=\mathfrak{D}\big(\{\brE(\brs,\tau)\}_{\tau\in T_e^d(t)},\{\brB(\brs,\tau)\}_{\tau\in T_b^d(t)},\brs,t\big) \nonumber\\
        \brH(\brs,t)=\mathfrak{H}\big(\{\brE(\brs,\tau)\}_{\tau\in T_e^h(t)},\{\brB(\brs,\tau)\}_{\tau\in T_b^h(t)},\brs,t\big)\nonumber.
\end{numcases}

\item A medium is nonbianisotropic if

\begin{numcases}{\,}
    \partial_{\brB}\mathfrak{D}=0 \nonumber\\
    \partial_{\brE}\mathfrak{H}=0.\nonumber
\end{numcases}

In this case, we have

\begin{numcases}{\,}
        \brD(\brs,t)=\mathfrak{D}\big(\{\brE(\mbox{\boldmath$\sigma$},\tau)\}_{(\mbox{\boldmath$\sigma$},\tau)\in S_e^d(\brs)\times T_e^d(t)},\brs,t\big) \nonumber\\
        \brH(\brs,t)=\mathfrak{H}\big(\{\brB(\mbox{\boldmath$\sigma$},\tau)\}_{(\mbox{\boldmath$\sigma$},\tau)\in S_b^h(\brs)\times T_b^h(t)},\brs,t\big)\nonumber
\end{numcases}

\item A medium is homogeneous in time, or, as is commonly said, stationary, if the functionals do not explicitly depend on the time $t$:

\begin{numcases}{\,}
    \partial_t\mathfrak{D}=0 \nonumber\\
    \partial_t\mathfrak{H}=0.\nonumber
\end{numcases}

The constitutive relations then get:
\begin{numcases}{\,}
        \brD(\brs,t)=\mathfrak{D}\big(\{\brE(\mbox{\boldmath$\sigma$},\tau)\}_{(\mbox{\boldmath$\sigma$},\tau)\in S_e^d(\brs)\times T_e^d(t)},\{\brB(\mbox{\boldmath$\sigma$},\tau)\}_{(\mbox{\boldmath$\sigma$},\tau)\in S_b^d(\brs)\times T_b^d(t)},\brs\big) \nonumber\\
        \brH(\brs,t)=\mathfrak{H}\big(\{\brE(\mbox{\boldmath$\sigma$},\tau)\}_{(\mbox{\boldmath$\sigma$},\tau)\in
S_e^h(\brs)\times
T_e^h(t)},\{\brB(\mbox{\boldmath$\sigma$},\tau)\}_{(\mbox{\boldmath$\sigma$},\tau)\in
S_b^h(\brs)\times T_b^h(t)},\brs\big).\nonumber
\end{numcases}

We would like to stress that this hypothesis is, in a way, delicate,
when treating nonlinear optics. Indeed, for the nonlinear effects to
be important, the intensity of light has to be large. But if this
intensity is too large, then we cannot neglect the effects of the
electromagnetic field on the medium. Some effects like saturation,
damage, etc. do appear, and we leave the realm of stationary media.
\end{itemize}

The transcriptions of locality in time (commonly called
nondispersive) or homogeneity in space in the inductive functions
should be obvious. We can then define the (local) electromagnetic
vacuum: a vacuum holds at a point $\brs$ and a time $t$ if the
following equalities of vectors are satisfied:

\begin{numcases}{\,}
    \brD(\brs,t)=\eps_0\,\brE(\brs,t) \nonumber\\
    \brH(\brs,t)=1/\mu_0\,\brB(\brs,t).\nonumber
\end{numcases}
This leads to define the `answer' of a medium to an
electromagnetic perturbation as the difference between the induction
in that medium and the one in a vacuum:

\begin{numcases}{\,}
    \brP_e:=\brD-\eps_0\,\brE \nonumber\\
    \brP_m:=\brH-1/\mu_0\,\brB. \nonumber
\end{numcases}
These two vector fields are called the electric polarization vector
field and the magnetic polarization vector field - we prefer to
write $\brP_m$ for what is usually denoted by $\brM$ (termed the
magnetization vector field) for the coherence of this article. We
now arrive at a key step. In the following definition, we denote by
$V'$ the dual of the vector space $V$.

\newtheorem{defONL}{Definition}
\begin{defONL}{Smooth Medium}\label{DefSmooth}

A nonbianisotropic medium is called smooth in a neighborhood of a
point $\brs$ and a time $t$ if the electric and magnetic
polarization vector fields admit Taylor expansions of the following
kind:

\begin{numcases}{\,}
    \brP_e=\brP_e^{(0)}+\sum_{n\in\N}\brP_e^{(n)} \nonumber\\
    \brP_m=\brP_m^{(0)}+\sum_{n\in\N}\brP_m^{(n)}\nonumber
\end{numcases}
with

\begin{numcases}{\,}
    \brP_e^{(n)}(\brs,t)=\eps_0\int_{-\infty}^{\infty}dt_1\cdots\int_{-\infty}^{\infty}dt_n\,Q_e^{(n)}(\brs,t;\brs_1,t_1,\cdots,\brs_n,t_n) \nonumber\\
    \qquad\brE(\brs_1,t_1)\cdots\brE(\brs_n,t_n) \nonumber\\
    \brP_m^{(n)}(\brs,t)=\eps_0\int_{-\infty}^{\infty}dt_1\cdots\int_{-\infty}^{\infty}dt_n\,Q_m^{(n)}(\brs,t;\brs_1,t_1,\cdots,\brs_n,t_n) \nonumber\\
    \qquad\brH(\brs_1,t_1)\cdots\brH(\brs_n,t_n)\nonumber
\end{numcases}
where

\begin{numcases}{\,}
Q_e^{(n)}:\R^4\times\R^{4n}\rightarrow\R^3\otimes(\R^3)'^{\otimes n} \nonumber\\
Q_m^{(n)}:\R^4\times\R^{4n}\rightarrow\R^3\otimes(\R^3)'^{\otimes
n}.\nonumber
\end{numcases}
\end{defONL}

A similar definition can be applied to bianisotropic media as well:
the tensors $Q_e^{(n)}$ and $Q_m^{(n)}$ have then to be contracted
with the $2^n$ combinations of $\brE$ and $\brB$. Though treating
these media, among which we find the chiral ones, do not bring
important new difficulties, we won't anymore be concerned with them,
to keep formulae of `reasonable' size.

Due to the nonbianisotropy hypothesis, the electric and magnetic
inductions can be treated separately. We will from now on
concentrate on the electric one. Hence, when no confusion is
possible, we will write $\brP$ in place of $\brP_e$, $Q^{(n)}$ is
place of $Q^{(n)}_e$, say the polarization vector instead of the
electric polarization vector, etc. The treatment required for the
magnetic part is completely similar.

The set of operators $Q^{(n)}$, when applied to
$(\brs,t;\brs_1,t_1,\cdots,\brs_n,t_n)$, give the effect, at the
point $\brs$ and time $t$, that the electric field, evaluated at the
points $\brs_1,\cdots,\brs_n$ and times $t_1,\cdots,t_n$ have on the
electric polarization vector at the point $\brs$ and time $t$. The
zero-th order of the electric polarization vector
$\brP^{(0)}(\brs,t)$ corresponds to a spontaneous nonzero electric
moment. It does not vanish for ferroelectric materials. For
convenience, we write $\brP^{(0)}(\brs,t)=\eps_0Q^{(0)}(\brs,t)$.
The term $\brP^{(n)}$, for $n\in\N$ is called the $n$-th order of
the polarization vector.

\subsection{A tractable expression for the polarization vector}

\subsubsection{The definition of the electric susceptibility tensors}

The expansion of the (electric) polarization vector given in the
definition (\ref{DefSmooth}) is still too complicated, and needs
further simplification to be studied. The following hypothesis,
still motivated for the presentation, is that the medium is
spatially local. The electric inductive functional has the form

$$\brD(\brs,t)=\mathfrak{D}\big(\{\brE(\brs,\tau)\}_{\tau\in
T_e^d(t)},\brs,t\big)$$
so that

\begin{align}\label{eqPnQ}
&\brP(\brs,t)=\brP^{(0)}(\brs,t)+\sum_{n\in\N}\eps_0\int_{-\infty}^{\infty}dt_1\cdots\int_{-\infty}^{\infty}dt_n\,Q^{(n)}(\brs,t;t_1,\cdots,t_n)\nonumber \\
&\qquad\brE(\brs,t_1)\cdots \brE(\brs,t_n).
\end{align}
Once again, we have adapted the functionals without changing the
name (now, we have
$Q^{(n)}:\R^4\times\R^{n}\rightarrow\R^3\otimes(\R^3)'^{\otimes
n}$).

Of more fundamental nature, let us suppose that the medium is
stationary; this means that, first, the spontaneous polarization
does not depend on time, and secondly, that for any point $\brs$ and
for any duration time $T$, the effect of the electric vector field
evaluated at the point $\brs$ and at the times $t_1, \cdots, t_n$ on
the polarization vector field evaluated at the point $\brs$ and at
the time $t$ is the same than the effect of the electric vector
field evaluated at the point $\brs$ and at the times $t_1-T, \cdots,
t_n-T$ on the polarization vector field evaluated at the point
$\brs$ and at the time $t-T$, i.e.

$$Q^{(0)}(\brs,t)=Q^{(0)}(\brs,t-T)$$
and

$$Q^{(n)}(\brs,t;t_1,\cdots,t_n)=Q^{(n)}(\brs,t-T;t_1-T,\cdots,t_n-T),\quad\forall n\in\N,\,\forall T\in\R.$$

When considering stationary media, a usual trick (\cite{Boyd}) is to
define

\begin{displaymath}
  \begin{array}{rlcl}
  R^{(0)} :  & \R^3 & \rightarrow & \R^3, \\
  & \brs & \mapsto & \displaystyle{Q^{(0)}(\brs,0)}, \\
  \end{array}
\end{displaymath}

\begin{displaymath}
  \begin{array}{rlcl}
  R^{(n)} :  & \R^3\times\R^n & \rightarrow & \R^3\otimes(\R^3)'^{\otimes n}, \\
  & (\brs,t_1,\cdots,t_n) & \mapsto & \displaystyle{Q^{(n)}(\brs,0;-t_1,\cdots,-t_n),\quad n\in\N}. \\
  \end{array}
\end{displaymath}
This function $R^{(n)}$ is called the response function of the
$n$-th order, because it allows to write the relation (\ref{eqPnQ})
as a convolution on $\R^n$ between $R^{(n)}$ and
$$\brE^{(n)}:\R^3\times\R^n\rightarrow(\R^3)^{\otimes
n},(\brs,t_1,\cdots,t_n)\mapsto\brE(\brs,t_1)\cdots \brE(\brs,t_n)
\,:$$

$$\brP^{(n)}(\brs,t)=\eps_0\big(R^{(n)}(\brs,\cdot)\ast_n\brE^{(n)}(\brs,\cdot)\big)(t)$$
$$=\eps_0\int_{-\infty}^{\infty}dt_1\cdots\int_{-\infty}^{\infty}dt_n\,R^{(n)}(\brs,t-t_1,\cdots,t-t_n)\brE(\brs,t_1)\cdots
\brE(\brs,t_n)$$%
In the last equation, we make the transformation $t_j\mapsto t-t_j$
to obtain

$$\brP^{(n)}(\brs,t)=\eps_0\int_{-\infty}^{\infty}dt_1\cdots\int_{-\infty}^{\infty}dt_n\,R^{(n)}(\brs,t_1,\cdots,t_n)\brE(\brs,t-t_1)\cdots
\brE(\brs,t-t_n)$$%
and then we express the electric field in its Fourier basis,

$$\brE(\brs,t-t_j)=\int_{-\infty}^{\infty}d\omega_j\,e^{-i\omega_j(t-t_j)}\hat{\brE}(\brs,\omega_j),\quad j\in\{1,\cdots,n\}.$$
This leads to

\begin{align}\label{Pnchi}
\brP^{(n)}(\brs,t)&=\eps_0\int_{-\infty}^{\infty}d\omega_1\,e^{-i\omega_1t}\cdots\int_{-\infty}^{\infty}d\omega_n\,e^{-i\omega_nt}\nonumber \\
&\qquad\qquad\underline{\chi}^{(n)}(\brs,\omega_1,\cdots,\omega_n)\hat{\brE}(\brs,\omega_1)\cdots\hat{\brE}(\brs,\omega_n)\nonumber \\
&=\eps_0\int_{-\infty}^{\infty}d\omega_1\cdots\int_{-\infty}^{\infty}d\omega_n\,\underline{\chi}^{(n)}(\brs,\omega_1,\cdots,\omega_n)\nonumber \\
&\qquad\qquad\hat{\brE}(\brs,\omega_1)\cdots\hat{\brE}(\brs,\omega_n)e^{-i(\omega_1+\cdots+\omega_n)t}
\end{align}
where

\begin{eqnarray}\label{DefUnderlineChin}
  \underline{\chi}^{(n)} : \R^3\times\R^n & \rightarrow & \C^3\otimes(\C^3)'^{\otimes n}, \\
  (\brs,\omega_1,\cdots,\omega_n) & \mapsto &
  \displaystyle{\int_{-\infty}^{\infty}dt_1\cdots\int_{-\infty}^{\infty}dt_n\,R^{(n)}(\brs,t_1,\cdots,t_n)e^{i\mbox{\boldmath$\omega$}\cdot\mathbf{t}}}\nonumber
\end{eqnarray}
with $\mbox{\boldmath$\omega$}=(\omega_1,\cdots,\omega_n)$,
$\mathbf{t}=(t_1,\cdots,t_n)$ and the usual duality product on
$\R^n$. We see that $\underline{\chi}^{(n)}$, which is a tensor
field of rank $n+1$, called the (electric) susceptibility tensor of
the $n$-th order, is the Fourier transform (up to a scalar
multiplication by $(2\pi)^n$) of the response function $R^{(n)}$ :

$$\underline{\chi}^{(n)}(\brs,\omega_1,\cdots,\omega_n)=(2\pi)^n\,\hat{R}^{(n)}(\brs,\omega_1,\cdots,\omega_n).$$

$\underline{\chi}^{(n)}(\brs,\omega_1,\cdots,\omega_n)$ gives the
way the components, evaluated at the point $\brs$, of the electric
field oscillating with the angular frequencies $\omega_1$, \dots,
$\omega_n$ contribute to the polarization vector oscillating at the
angular frequency $\omega_1+\cdots+\omega_n$ at the same place
$\brs$. The unit of $\underline{\chi}^{(n)}$ is $m^{n-1}\,V^{1-n}$.

By symmetry, we define $\underline{\chi}^{(0)}$ by

\begin{displaymath}
  \begin{array}{rlcl}
  \underline{\chi}^{(0)} :  & \R^3 & \rightarrow & \R^3, \\
  & \brs & \mapsto & \displaystyle{R^{(0)}(\brs)}. \\
  \end{array}
\end{displaymath}

\subsubsection{The symmetries of the
susceptibility tensors}

The susceptibility tensor fields exhibit several symmetries. These
ones are important to better understand these tensors, and will
allow to greatly simplify the results of the following sections,
about propagation equation systems and energy criteria.

\paragraph{The intrinsic permutation symmetry}

We will now present the intrinsic permutation symmetry of the
susceptibility tensors. Although the result of this paragraph is
well-known (\cite{bBloembergen65,Boyd,Jonsson}), we demonstrate it
because we think that the way we derive it is more systematic. For
this, let us develop the equation (\ref{Pnchi}) in components:

$$P^{(n)\,i}(\brs,t)=\eps_0\int_{-\infty}^{\infty}d\omega_1\cdots\int_{-\infty}^{\infty}d\omega_n\,\underline{\chi}^{(n)\,i}_{\qquad i_1\,\cdots\,i_n}(\brs,\omega_1,\cdots,\omega_n)$$
$$\hat{E}^{i_1}(\brs,\omega_1)\cdots\hat{E}^{i_n}(\brs,\omega_n)e^{-i(\omega_1+\cdots+\omega_n)t},$$
where the summation convention over repeated indices is used. The
intrinsic permutation symmetry is based on the fact that the
$\underline{\chi}^{(n)\,i}_{\qquad i_1\,\cdots\,i_n}$ are not
uniquely defined: for $j$ in $\{1,\cdots,n\}$, the indexes $i_j$ and
the variables $\omega_j$ are dummy, so for any bijection $\sigma$
from $\{1,\cdots,n\}$ to itself, we have:

$$P^{(n)\,i}(\brs,t)=\eps_0\int_{-\infty}^{\infty}d\omega_{\sigma 1}\cdots\int_{-\infty}^{\infty}d\omega_{\sigma n}\,\underline{\chi}^{(n)\,i}_{\qquad i_{\sigma 1}\,\cdots\,i_{\sigma n}}(\brs,\omega_{\sigma 1},\cdots,\omega_{\sigma n})$$
$$\hat{E}^{i_{\sigma 1}}(\brs,\omega_{\sigma 1})\cdots\hat{E}^{i_{\sigma n}}(\brs,\omega_{\sigma n})e^{-i(\omega_{\sigma 1}+\cdots+\omega_{\sigma n})t}.$$
Thus, denoting by $\mathcal{S}_n$ the symmetric group on a set of
cardinality $n$ (we recall that this group is of order $n!$), we
have

\begin{displaymath}
P^{(n)\,i}(\brs,t)=\frac{\eps_0}{n!}\sum_{\sigma\in\mathcal{S}_n}
\int_{-\infty}^{\infty}d\omega_{\sigma 1}
\cdots\int_{-\infty}^{\infty}d\omega_{\sigma n}\,
\underline{\chi}^{(n)\,i}_{\qquad i_{\sigma 1}\,\cdots\,i_{\sigma
n}} (\brs,\omega_{\sigma 1},\cdots,\omega_{\sigma n})
\end{displaymath}
\begin{displaymath}
\hat{E}^{i_{\sigma 1}}(\brs,\omega_{\sigma 1})
\cdots\hat{E}^{i_{\sigma n}}(\brs,\omega_{\sigma n})
e^{-i(\omega_{\sigma 1}+\cdots+\omega_{\sigma n})t}.
\end{displaymath}
But, loosely speaking, $\int_{-\infty}^{\infty}d\omega_{\sigma 1}
\cdots\int_{-\infty}^{\infty}d\omega_{\sigma
n}=\int_{-\infty}^{\infty}d\omega_1\cdots\int_{-\infty}^{\infty}d\omega_n$
and

\begin{displaymath}\prod_{j=1}^n\hat{E}^{i_{\sigma
j}}(\brs,\omega_{\sigma j})e^{-i\omega_{\sigma j}t}=\prod_{j=1}^n
\hat{E}^{i_j}(\brs,\omega_j)e^{-i\omega_jt}
\end{displaymath}
for any $\sigma$ in $\mathcal{S}_n$. We thus have

\begin{displaymath}
P^{(n)\,i}(\brs,t)=\eps_0%
\int_{-\infty}^{\infty}d\omega_1
\cdots\int_{-\infty}^{\infty}d\omega_n\,%
\{\frac{1}{n!}\sum_{\sigma\in\mathcal{S}_n}\,
\underline{\chi}^{(n)\,i}_{\qquad i_{\sigma 1}\,\cdots\,i_{\sigma
n}} (\brs,\omega_{\sigma 1},\cdots,\omega_{\sigma n})\}
\end{displaymath}
\begin{displaymath}
\hat{E}^{i_1}(\brs,\omega_1) \cdots\hat{E}^{i_n}(\brs,\omega_n)
e^{-i(\omega_1+\cdots+\omega_n)t}.
\end{displaymath}

This leads to define the (not underlined) tensor
$\chi^{(n)}=\bre_i\,\chi^{(n)\,i}_{\qquad
i_1\cdots\,i_n}\otimes\bre^{i_1}\otimes\cdots\otimes\bre^{i_n}$,
once again called susceptibility tensor of order $n$, by

\begin{eqnarray}\label{DefChin}
  \chi^{(n)} : \R^3\times\R^n & \rightarrow & \C^3\otimes(\C^3)'^{\otimes n}, \\
  (\brs,\omega_1,\cdots,\omega_n) & \mapsto & \frac{1}{n!}\sum_{\sigma\in\mathcal{S}_n}\bre_i\,\underline{\chi}^{(n)\,i}_{\qquad i_{\sigma 1}\,\cdots\,i_{\sigma n}} (\brs,\omega_{\sigma 1},\cdots,\omega_{\sigma n}) \nonumber\\
  & & \qquad\qquad\qquad\otimes\bre^{i_1}\otimes\cdots\otimes\bre^{i_n}, \nonumber
\end{eqnarray}
and to deduce the following expression of the polarization vector of
the $n$-th order:

\begin{equation}\label{Pnchisympermintr}
\brP^{(n)}(\brs,t)=\eps_0%
\int_{-\infty}^{\infty}d\omega_1
\cdots\int_{-\infty}^{\infty}d\omega_n\,\chi^{(n)}
(\brs,\omega_1,\cdots,\omega_n)
\end{equation}

\begin{equation*}
\hat{\brE}(\brs,\omega_1)\cdots\hat{\brE}(\brs,\omega_n)
e^{-i(\omega_1+\cdots+\omega_n)t},
\end{equation*}
with a symmetric $\chi^{(n)}$ in the sense that

$$\chi^{(n)\,i}_{\qquad i_{\tau 1}\,\cdots\,i_{\tau n}}
(\brs,\omega_{\tau 1},\cdots,\omega_{\tau n})=\chi^{(n)\,i}_{\qquad
i_1\,\cdots\,i_n}
(\brs,\omega_1,\cdots,\omega_n),\quad\forall\tau\in\mathcal{S}_n.$$

This symmetry of the (not underlined) susceptibility tensor is
called the intrinsic permutation symmetry. It allows for example to
consider only sets of angular frequencies
$(\omega_1,\cdots,\omega_n)$ ordered such that
$\omega_1\leqslant\omega_2\leqslant\cdots\leqslant\omega_n$.

Finally, since the zero-th order susceptibility tensor does not
present degeneracy, we define $\chi^{(0)}:=\underline{\chi}^{(0)}$.

%\paragraph{An other degeneracy}
%
%If $\omega_j=\omega_{j+1}$ for some $j$ in $\{1,\cdots,n-1\}$, an
%other degeneracy appears:
%
%$$\chi^{(n)\,i}_{\qquad i_1\,\cdots\,i_j\,i_{j+1}\,\cdots\,i_n}
%(\brs,\omega_1,\cdots,\omega_j,\omega_j,\cdots,\omega_n),$$%
%when contracted with the electric vectors
%$\hat{\brE}(\brs,\omega_1)$, $\cdots$, $\hat{\brE}(\brs,\omega_n)$,
%give the same result than $\chi^{(n)\,i}_{\qquad
%i_1\,\cdots\,i_{j+1}\,i_j\,\cdots\,i_n}
%(\brs,\omega_1,\cdots,\omega_j,\omega_j,\cdots,\omega_n)$ when
%contracted with the same electric vectors. This leads to define a
%new tensor, so that its components are uniquely defined. In our
%case, we have to replace
%
%$$\chi^{(n)\,i}_{\qquad i_1\,\cdots\,i_j\,i_{j+1}\,\cdots\,i_n}
%(\brs,\omega_1,\cdots,\omega_j,\omega_j,\cdots,\omega_n)$$
%by
%
%$$\frac{1}{2}\Big(\chi^{(n)\,i}_{\qquad i_1\,\cdots\,i_j\,i_{j+1}\,\cdots\,i_n}
%(\brs,\omega_1,\cdots,\omega_j,\omega_j,\cdots,\omega_n)$$
%$$\qquad+\chi^{(n)\,i}_{\qquad i_1\,\cdots\,i_{j+1}\,i_j\,\cdots\,i_n}
%(\brs,\omega_1,\cdots,\omega_j,\omega_j,\cdots,\omega_n)\Big).$$%
%We won't go further in this study and report the interested reader
%to \cite{Jonsson,bButcher91}.\textbf{ It may seem strange that we do
%for these degeneracies exactly what we reproach to the others when
%they treat the intrinsic permutation symmetry\dots Also, we did not
%apply this, so I am not sure that the $\chi^{(3)}$ coefficient we
%choose match the one given in the literature or by the
%companies\dots}

\vspace{.2cm}

\paragraph{The Hermitian symmetry}\label{parHermSym}

The electric field (and the polarization vector field) $\brE$ being
real, the harmonic components satisfy
$\hat{\brE}(\brs,\omega)=\overline{\hat{\brE}}(\brs,-\omega)$ for
all $\brs$ in $\R^3$. This leads to the Hermitian symmetry of the
susceptibility tensor fields:

$$\overline{\chi^{(n)}}(\brs,\omega_1,\cdots,\omega_n)=\chi^{(n)}(\brs,-\omega_1,\cdots,-\omega_n).$$

\section{The propagation equation systems}

\subsection{The general propagation equation
system}\label{SubsecGenProgEqSyst}

\paragraph{The magnetic response}

The preceding section shows how the electric answer of a medium is
encoded in the (electric) susceptibility tensors. A similar
treatment shows how the magnetic answer of a medium is encoded in
the magnetic susceptibility tensors. Nevertheless, for the sake of
simplicity of this paper, and also because at the wavelength we are
looking at, nonlinear magnetic effects are usually negligible, we
suppose from now on that the magnetic characteristics of the media
are smooth, linear, local in space and stationary. This means that
the magnetic inductive functional has the form

$$\brH(\brs,t)=\mathfrak{H}\big(\{\brB(\brs,\tau)\}_{\tau\in
T_b^h(t)},\brs\big)$$ so that

\begin{equation*}%\label{eqPnQ}
\brP_m(\brs,t)=\mu_0^{-1}\int_{-\infty}^{\infty}d\omega\,\chi_m^{(1)}(\brs,\omega)\hat{\brB}(\brs,\omega)e^{-i\omega
t}.
\end{equation*}
To recover the usual conventions, we define the relative
permittivity by:

$$\hat{\mu}_r(\brs,\omega)=(1+\chi_m^{(1)}(\brs,\omega))^{-1},$$
so that

\begin{equation}\label{eqBH}
\hat{\brB}=\hat{\mu}\hat{\brH}.
\end{equation}

This last equation closes the sets of Maxwell's equations. Reporting
the expressions of the inductive vector fields in terms of the
primitive fields ((\ref{Pnchisympermintr}) as well as the
development of the polarization vector field in the definition
\ref{DefSmooth} for the electric response, (\ref{eqBH}) for the
magnetic response) in the first set of Maxwell's equations leads to
the propagation equations. This section is devoted to present them,
and to suggest some approximations in order to be able to implement
them in a computer program.

\paragraph{The general propagation equation system}

On physical grounds, we suppose that the permeability tensor never
vanishes so that the two vectorial Maxwell's equations lead to

\begin{equation}\label{progED}
\nabla\times\big(\hat{\mu}^{-1}(\brs,\omega)\nabla\times\hat{\brE}(\brs,\omega)\big)-\omega^2\hat{\brD}(\brs,\omega)=i\omega\hat{\brJ}(\brs,\omega)
\end{equation}
for all $\omega$ in $\R$ and $\brs$ in $\R^3$.

The dependence of $\brD$ in $\brE$ and in the (electric)
characteristic of the medium were obtained in the first section, so
that we just have to insert them in this expression. For this we
have to write the polarization vector in the Fourier basis:

\begin{align*}
\hat{\brP}^{(n)}(\brs,\omega)&=\frac{1}{2\pi}\int_{-\infty}^{\infty}dt\,\brP^{(n)}(\brs,t)e^{i\omega
t} \\
&=\frac{\eps_0}{2\pi}\int_{-\infty}^{\infty}dt\int_{-\infty}^{\infty}d\omega_1
\cdots\int_{-\infty}^{\infty}d\omega_n\,\chi^{(n)}
(\brs,\omega_1,\cdots,\omega_n) \\
&\quad \hat{\brE}(\brs,\omega_1)\cdots\hat{\brE}(\brs,\omega_n)
e^{-i(\omega_1+\cdots+\omega_n-\omega)t}, \\
&=\eps_0\int_{-\infty}^{\infty}d\omega_1
\cdots\int_{-\infty}^{\infty}d\omega_n\,\chi^{(n)}
(\brs,\omega_1,\cdots,\omega_n) \\
&\quad \hat{\brE}(\brs,\omega_1)\cdots\hat{\brE}(\brs,\omega_n)
\delta(\omega_1+\cdots+\omega_n-\omega)
\end{align*}
where $\delta$ is the Dirac distribution; the last equation follows
upon integration upon the time variable. For the spontaneous
polarization vector, we have

$$\hat{\brP}^{(0)}(\brs,\omega)=\eps_0\chi^{(0)}(\brs)\delta(\omega).$$

The propagation equation system is therefore

\begin{equation}\label{EqProgWithoutNot}
\nabla\times\big(\hat{\mu}^{-1}(\brs,\omega)\nabla\times\hat{\brE}(\brs,\omega)\big)
-\omega^2\eps_0\Big(\hat{\brE}(\brs,\omega)+\chi^{(0)}(\brs)\delta(\omega)
\end{equation}
$$+\sum_{n\in\N}\int_{-\infty}^{\infty}d\omega_1
\cdots\int_{-\infty}^{\infty}d\omega_n\,\chi^{(n)}
(\brs,\omega_1,\cdots,\omega_n)
\hat{\brE}(\brs,\omega_1)\cdots\hat{\brE}(\brs,\omega_n)$$
$$\delta(\omega_1+\cdots+\omega_n-\omega)\Big)=i\omega\hat{\brJ}(\brs,\omega).$$

\subsection{Nonlinearity of the second order}\label{SubsecNLorder2}

\paragraph{The introduction of some notations}

Of course, the propagation equation system (\ref{EqProgWithoutNot})
is too complicated to be solved directly. We will thus suppose from
now on that no nonlinearity higher than the quadratic one are
present. We will treat this order of nonlinearity with some detail;
the other ones being direct but more and more sophisticated
generalizations of this case. Also, though its presence does not
lead to high difficulties, we will neglect the spontaneous
polarization. The propagation equation system thus becomes

$$\nabla\times\big(\hat{\mu}^{-1}(\brs,\omega)\nabla\times\hat{\brE}(\brs,\omega)\big)$$
$$-\omega^2\eps_0\Big(\hat{\brE}(\brs,\omega)
+\int_{-\infty}^{\infty}d\omega_1\,\chi^{(1)}
(\brs,\omega_1)\hat{\brE}(\brs,\omega_1) \delta(\omega_1-\omega)$$
$$+\int_{-\infty}^{\infty}d\omega_1\int_{-\infty}^{\infty}d\omega_2\,\chi^{(2)}
(\brs,\omega_1,\omega_2)
\hat{\brE}(\brs,\omega_1)\hat{\brE}(\brs,\omega_2)
\delta(\omega_1+\omega_2-\omega)\Big)$$
$$=i\omega\hat{\brJ}(\brs,\omega)$$
for all $\omega$ in $\R$ and $\brs$ in $\R^3$.

In order to facilitate the reading of these equations, we introduce
some notations. First, the usual linear permittivity:
$\widehat{\eps_r}^{(1)}(\brs,\omega):=1+\chi^{(1)}(\brs,\omega)$.
Using this and integrating the Dirac distributions, we have

$$\nabla\times\big(\hat{\mu}^{-1}(\brs,\omega)\nabla\times\hat{\brE}(\brs,\omega)\big)-\omega^2\eps_0\Big(\widehat{\eps_r}^{(1)}(\brs,\omega)\hat{\brE}(\brs,\omega)$$
$$+\int_{-\infty}^{\infty}d\omega_1\,\chi^{(2)}(\brs,\omega_1,\omega-\omega_1)
\hat{\brE}(\brs,\omega_1)\hat{\brE}(\brs,\omega-\omega_1)\Big)=i\omega\hat{\brJ}(\brs,\omega)$$
for all $\omega$ in $\R$ and $\brs$ in $\R^3$.

Then, we define the linear Maxwell's operator by

$$\mathcal{M}_{(\brs,\omega)}^{lin}\big(\hat{\brE}(\brs,\omega)\big):=-\eps_0^{-1}\nabla\times\big(\hat{\mu}^{-1}(\brs,\omega)\nabla\times\hat{\brE}(\brs,\omega)\big)
+\omega^2\widehat{\eps_r}^{(1)}(\brs,\omega)\hat{\brE}(\brs,\omega);$$%
this operator is defined such that, in a linear media, we have

$$\mathcal{M}_{(\brs,\omega)}^{lin}\big(\hat{\brE}(\brs,\omega)\big)=\frac{-i\omega}{\eps_0}\hat{\brJ}(\brs,\omega)$$
for all $\omega$ in $\R$ and $\brs$ in $\R^3$. Finally, we will use
the notation

$$\dfl\hat{\brE}(\brs,\omega_1),\cdots,\hat{\brE}(\brs,\omega_n)\ffl:=\chi^{(n)}(\brs,\omega_1,\cdots,\omega_n)\hat{\brE}(\brs,\omega_1)\cdots\hat{\brE}(\brs,\omega_n);$$
this term is the contribution to the polarization vector
$\hat{\brP}(\brs,\omega_1+\cdots+\omega_n)$ of the interaction
between the vectors $\hat{\brE}(\brs,\omega_1)$, $\cdots$,
$\hat{\brE}(\brs,\omega_n)$. The propagation equation system is now

\begin{equation}\label{EqProgWithoutHarmHyp}
\mathcal{M}_{(\brs,\omega)}^{lin}\big(\hat{\brE}(\brs,\omega)\big)+\omega^2\int_{-\infty}^{\infty}d\omega_1\,\dfl\hat{\brE}(\brs,\omega_1),\hat{\brE}(\brs,\omega-\omega_1)\ffl=\frac{-i\omega}{\eps_0}\hat{\brJ}(\brs,\omega)
\end{equation}
for all $\omega$ in $\R$ and $\brs$ in $\R^3$.

\paragraph{The harmonic assumption and the definition of the degree}

We focus on the case where the incident vector field is
monochromatic:
$\brE^i(\brs,t)=\brE_1^i(\brs)e^{-i\omega_It}+\brE_{-1}^i(\brs)e^{i\omega_It}$
(we note $\brE_p^i(\brs)$ for $\hat{\brE}^i(\brs,p\omega_I)$). To
simplify the system (\ref{EqProgWithoutHarmHyp}), which is described
by a continuous parameter, we suppose that the susceptibility tensor
field is nonzero only when it is evaluated on the harmonics of the
incident angular frequency

$$\chi^{(n)}(\brs,\omega_1,\cdots,\omega_n)=0$$%
if there is a $j$ in $\{1,\cdots,n\}$ such that no integer $p$
satisfies $\omega_j=p\omega_I$. This is an important hypothesis,
which implies that we will concentrate on the harmonic generations
and neglect the Raman or Brilloin scatterings, or subharmonic
generations. On one hand, we have to admit that this is a serious
stretch to the general aim stated in the introduction. On the other
hand, the set of harmonic generations is sufficiently important and
large to pursue this study.

The propagation equation of the $p$-th harmonic is now (we stop to
explicitly write the point $\brs$ where the fields are evaluated -
note that no confusion can appear, because of the locality of the
considered media; this is just a return to equalities between vector
fields on $\R^3$ and not simply vectors in $\R^3$ - we also write
$\mathcal{M}_p^{lin}$ for $\mathcal{M}_{(\brs,p\omega_I)}^{lin}$)

\begin{equation}\label{EqProgEn2}
\mathcal{M}_p^{lin}\big(\brE_p\big)+(p\omega_I)^2\sum_{q\in\Z}\,\dfl\brE_q,\brE_{p-q}\ffl=\frac{-i\,p\omega_I}{\eps_0}\brJ_p,
\end{equation}
for all $p$ in $\Z$. Since the sources oscillates only at the
angular frequency $\omega_I$ (and of course $-\omega_I$), we have
$\brJ_p=\brJ_p\delta_{|p|,1}$. We note that the harmonic assumption
turns the propagation equation system from a system described by a
parameter ranging on a \emph{continuous} set to a system described
by a parameter ranging on a \emph{discrete} set.

Nevertheless, (\ref{EqProgEn2}) is still too complicated to be
solved: it still contains an \emph{infinite} number of equations. To
this aim, we introduce the notion of the \textbf{degree}, whose
effect will be to obtain propagation equation systems described by a
parameter ranging on a \emph{finite} set. No absolute definition of
the degree does exist; indeed, the way we will simplify the system
(\ref{EqProgEn2}) will depend on our purpose. To illustrate the
choice we have to settle on, we give two definitions, and will argue
that these two definitions present interesting, but different points
of view.

\begin{defONL}{the degree $d_1$ of the approximation at the $n$-th order
of a monochromatic field is}\label{DefDegre1}

\begin{equation*}
d_1:=\min_{\underline{d}\in\N}\{||(p_1,\cdots,p_n)||_{l_1(\Z)}>\underline{d}\Rightarrow\chi^{(n)}(\brs;p_1\omega_I,\cdots,p_n\omega_I)=0,\forall\brs\in\R^3\}
\end{equation*}
\end{defONL}
\begin{defONL}{the degree $d_{\infty}$ of the approximation at the $n$-th order of
a monochromatic field is}\label{DefDegreinfty}

\begin{align*}
d_{\infty}&:=\min_{\underline{d}\in\N}\{||(p_1,\cdots,p_n,p_1+\cdots+p_n)||_{l_{\infty}(\Z)}>\underline{d} \\
&\qquad\Rightarrow\chi^{(n)}(\brs;p_1\omega_I,\cdots,p_n\omega_I)=0,\forall\brs\in\R^3\}
\end{align*}
\end{defONL}
With words, this means that, since we can consider only a finite
number of components of the electric field, we suppose that the
susceptibility tensors vanish if they are evaluated at high enough
frequencies. In the $d_1$-th degree $d$, all the terms of the kind
$\dfl\brE_{p_1},\cdots,\brE_{p_n}\ffl$ vanish as soon as
$|p_1|+\cdots+|p_n|$ is strictly greater than $d$. In the
$d_{\infty}$-th degree $d$, all the terms of the kind
$\dfl\brE_{p_1},\cdots,\brE_{p_n}\ffl$ vanish as soon as $|p_1|$,
$\cdots$, $|p_n|$ or $|p_1+\cdots+p_n|$ is strictly greater than
$d$. In both cases, the electric vector can be written, for all
$\brs$ in $\R^3$ and all $t$ in $\R$ as

\begin{align*}
\brE(\brs,t)&=\sum_{\substack{p\in\Z
\\ |p|\leqslant d}}\brE_p(\brs)e^{-ip\omega_It} \\
&=\brE_0+2\RE\{\sum_{1\leqslant p\leqslant
d}\brE_p(\brs)e^{-ip\omega_It}\},
\end{align*}
since $\brE_{-p}=\overline{\brE_p}$.

We hope that these definitions of the degree will become intuitive
when reading the next subsection.

\subsection{The propagation equation systems in the lowest degree}\label{SubsecSecOrder}
We add a new hypothesis: the static component of the electric field
$\brE_0$ vanishes. Keeping it would not have brought huge
difficulties, at least in this theoretical work. We note that this
prevents us from studying the Pockels effect or the optical
rectification.

\subsubsection{In the degree 1 (linear case)}
The cases $d_1=1$ and $d_{\infty}=1$ are both:

\begin{equation*}%\label{eqProgOrdre2Degre1}
\mathcal{M}_1^{lin}(\brE_1)=\frac{-i\,\omega_I}{\eps_0}\brJ_1.
\end{equation*}
This means that, whatever we use the $d_1$ or the $d_{\infty}$
degree, the coarser approximation that we can do in nonlinear optics
reduces to the linear case.

\vspace{.2cm}

\newtheorem{remONL}{Remark}
\begin{remONL}{The symmetry of the propagation equation system}%\label{DefSmooth}

Due to the Hermitian symmetry of the Fourier components of the
electric field ($\brE_{-p}=\overline{\brE_p}$), the one of the
susceptibility tensor (see the paragraph \emph{The Hermitian
symmetry}, page \pageref{parHermSym}) and the one of the
permeability tensor
($\mu(\brs,-\omega)=\overline{\mu(\brs,\omega)}$), it can be shown
that the system (\ref{EqProgEn2}) is coherent in the following
sense: if $\brE_p$ is a solution of the propagation equation for
$\brE_p$, then $\overline{\brE_p}$ is a solution of the propagation
equation for $\brE_{-p}$. This allows to give only the equations for
positive frequencies.
\end{remONL}

\subsubsection{In the degree 2 (second harmonic generation)}\label{subsubsectionGenerationSecondeHarmonique}

\paragraph{With the $d_1$ definition}
In the $d_1=2$ case, we have

\begin{subequations}\label{eqProgOrdre2Degre2}
\begin{align}
\mathcal{M}_1^{lin}(\brE_1)=\frac{-i\,\omega_I}{\eps_0}\brJ_1, \\
\nonumber \\
\mathcal{M}_2^{lin}(\brE_2)+(2\omega_I)^2\dfl\brE_1,\brE_1\ffl=0.
\end{align}
\end{subequations}

Indeed, in this degree $d_1=2$, $\chi^{(2)}$ can be contracted only
with the components $\brE_p$ and $\brE_q$ such that
$|p|+|q|\leqslant2$; since $\brE_0=0$, we also have $p\neq0$ and
$q\neq0$. Hence, the nonzero susceptibility tensors are
$\chi^{(2)}(-\omega_I,-\omega_I)$, $\chi^{(2)}(\omega_I,-\omega_I)$,
$\chi^{(2)}(-\omega_I,\omega_I)$ et $\chi^{(2)}(\omega_I,\omega_I)$.
But the second and the third ones do not appear in the propagation
equation system since we suppose that $\brE_0$ vanishes, and the
first one describes the generation of the component $\brE_{-2}$,
that we directly obtain from the equation that satisfies $\brE_2$.

This propagation equation system has the advantage that it can be
solved treating only linear equations: one first solves the equation
that $\brE_1$ satisfies and then the one that $\brE_2$ satisfies,
treating $\brE_1$ as a source. But this does not mean that the
system is linear: if the incident field is multiplied by a constant
factor $m$, $\brE_1$ is scaled by $m$ whereas $\brE_2$ is scaled by
$m^2$.

Since the equation that satisfies $\brE_1$ is linear, it is commonly
said that we are in the framework of the nondepletion of the pump
beam. Indeed, $\brE_1$ can be seen as a tank: the fact that it
generates $\brE_2$ does not change its energy.

\paragraph{With the $d_{\infty}$ definition}
In the $d_{\infty}=2$ case, we have

\begin{subequations}\label{eqProgOrdre2DegreI2}
\begin{align}
&\mathcal{M}_1^{lin}(\brE_1)+2\omega_I^2\dfl\brE_{-1},\brE_2\ffl=\frac{-i\,\omega_I}{\eps_0}\brJ_1, \\
\nonumber \\
&\mathcal{M}_2^{lin}(\brE_2)+(2\omega_I)^2\dfl\brE_1,\brE_1\ffl=0.
\end{align}
\end{subequations}

Considering this system, the definition of the degree $d_{\infty}$
can appear to be nonphysical: in the generic case, that is without
specifying a material, there are no reason for keeping the term
$\dfl\brE_{-1},\brE_2\ffl$ when neglecting $\dfl\brE_1,\brE_2\ffl$
that could generate the third harmonic (we recall that
$\|\brE_{-1}\|=\|\brE_1\|$). This remark argues in the sense that
the degree $d_1$ is the most natural one from the physical point of
view. Nevertheless, in practical cases, the degree $d_{\infty}$ can
be important since it allows to consider a lossless medium - a fact,
as seen above, impossible with the $d_1$ definition. The details are
given in the section \ref{SectionEnergy}

\subsubsection{In the degree 3 (third harmonic generation by a cascade effect)}\label{subsubsectionGenerationTroisiemeHarmonique}

A small number of interactions was considered in the degree $2$. In
the degree $3$, we consider the interactions that result from a
cascade effect. With the $d_1$-degree, the interactions between
$\brE_{-2}$ and $\brE_{2}$ on one side, and the components
$\brE_{-1}$ and $\brE_{1}$ on the other side are taken into account.
With the $d_{\infty}$-degree, we consider all the interactions where
$\brE_p$ occur for $p$ in $\{-3,-2,-1,1,2,3\}$.

\paragraph{With the $d_1$ definition}
In the $d_1=3$ case, we have

$$\mathcal{M}_1^{lin}(\brE_1)+\omega_I^2\Big(\dfl\brE_{-1},\brE_2\ffl+\dfl\brE_2,\brE_{-1}\ffl\Big)=\frac{-i\,\omega_I}{\eps_0}\brJ_1,$$

$$\mathcal{M}_2^{lin}(\brE_2)+(2\omega_I)^2\dfl\brE_1,\brE_1\ffl=0,$$

$$\mathcal{M}_3^{lin}(\brE_3)+(3\omega_I)^2\Big(\dfl\brE_1,\brE_2\ffl+\dfl\brE_2,\brE_1\ffl\Big)=0.$$
Using the intrinsic permutation symmetry, the propagation equation
system gets

\begin{subequations}%\label{eqProgOrdre2Degre3}
\begin{align*}
\mathcal{M}_1^{lin}(\brE_1)+2\omega_I^2\dfl\brE_{-1},\brE_2\ffl=\frac{-i\,\omega_I}{\eps_0}\brJ_1, \\
 \\
\mathcal{M}_2^{lin}(\brE_2)+(2\omega_I)^2\dfl\brE_1,\brE_1\ffl=0, \\
 \\
\mathcal{M}_3^{lin}(\brE_3)+2(3\omega_I)^2\dfl\brE_1,\brE_2\ffl=0. \\
\end{align*}
\end{subequations}

\paragraph{With the $d_{\infty}$ definition}
In the $d_{\infty}=3$ case, we have

\begin{subequations}
\begin{align*}
&\mathcal{M}_1^{lin}(\brE_1)+2\omega_I^2\Big(\dfl\brE_{-2},\brE_3\ffl+\dfl\brE_{-1},\brE_2\ffl\Big)=\frac{-i\,\omega_I}{\eps_0}\brJ_1, \\
 \\
&\mathcal{M}_2^{lin}(\brE_2)+(2\omega_I)^2\Big(2\dfl\brE_{-1},\brE_3\ffl+\dfl\brE_1,\brE_1\ffl\Big)=0, \\
 \\
&\mathcal{M}_3^{lin}(\brE_3)+2(3\omega_I)^2\dfl\brE_1,\brE_2\ffl=0.
\end{align*}
\end{subequations}

\vspace{.2cm}

It may seem curious that we consider the third harmonic generation
without taking into account the third order polarization. In fact,
we advise to the reader not to use these systems - they were exposed
only to give better intuition on the notion of degree.

In the second order of nonlinearity, the relevant systems are the
ones obtain in the degree $2$. It is a general fact that the $n$-th
harmonic generation is to be studied in the $n$-th order of
nonlinearity and in the $n$-th degree (at least with the two
definition of the degree presented here).

\subsection{Nonlinearity of the third order}\label{SubsecThirdOrder}

The remark given in the second order of nonlinearity and in the
third degree argues that we have to take into account the third
order polarization vector field to treat the third harmonic
generation. Roughly speaking, the study of the nonlinearity of the
second order was aimed for presenting our method. The `interesting'
propagation equation system, (\ref{eqProgOrdre2Degre2}) and
(\ref{eqProgOrdre2DegreI2}), show little interaction. In the third
order, more interactions are considered, so that the physical
effects are richer: the third harmonic generation, the optical
Kerr-effect, cascade effects. To study Raman scattering, one has to
generalize the result of this subsection to non-harmonic processes,
as in done in \cite{bGodard10}.

A treatment similar to the one used to obtain the system
(\ref{EqProgEn2}) has to be done with now
$\brP=\brP^{(1)}+\brP^{(2)}+\brP^{(3)}$. The propagation equation
system we obtain is

\begin{equation}\label{EqProgEn3}
\mathcal{M}_p^{lin}(\brE_p)+(p\omega_I)^2\Big(\sum_{q\in\Z}\dfl\brE_q,\brE_{p-q}\ffl+\sum_{(q,r)\in\Z^2}\dfl\brE_q,\brE_r,\brE_{p-q-r}\ffl\Big)=\frac{-i\,p\omega_I}{\eps_0}\brJ_p.
\end{equation}

In agreement with the last paragraph of the last subsection, the
interesting cases in the third order of nonlinearity are when the
degree is also equal to three. We will thus consider only these
cases. It is still assumed that the incident field is monochromatic,
so that $\brJ_p=\brJ_p\delta_{p,|1|}$.

\paragraph{With the $d_1$ definition}
In the $d_1=3$ case, we have

\begin{subequations}\label{eqProgOrdre3Degre3}
\begin{align}
\mathcal{M}_1^{lin}(\brE_1)+\omega_I^2\Big(2\dfl\brE_{-1},\brE_2\ffl+3\dfl\brE_{-1},\brE_1,\brE_1\ffl\Big)=\frac{-i\,\omega_I}{\eps_0}\brJ_1, \label{eqProgOrdre3Degre31} \\
\nonumber \\
\mathcal{M}_2^{lin}(\brE_2)+(2\omega_I)^2\dfl\brE_1,\brE_1\ffl=0, \label{eqProgOrdre3Degre32} \\
\nonumber \\
\mathcal{M}_3^{lin}(\brE_3)+(3\omega_I)^2\Big(2\dfl\brE_1,\brE_2\ffl+\dfl\brE_1,\brE_1,\brE_1\ffl\Big)=0. \label{eqProgOrdre3Degre33} \\
\nonumber\end{align}
\end{subequations}

This system presents several effects: the term
$\dfl\brE_{-1},\brE_2\ffl$ describes the counteraction of $\brE_2$
on $\brE_1$ (as it appears with the second order of nonlinearity and
the degree $d_{\infty}=2$), the pump beam is thus depleted;
$\dfl\brE_{-1},\brE_1,\brE_1\ffl$ describes the optical Kerr-effect;
lastly, the third harmonic is generated by two different processes:
the cascade effect $\dfl\brE_1,\brE_2\ffl$, and the third-order
process $\dfl\brE_1,\brE_1,\brE_1\ffl$.

\paragraph{With the $d_{\infty}$ definition}
In the $d_{\infty}=3$ case, we have

\begin{subequations}\label{SystemEqOrdre3Degre3E0nul}
\begin{align}
\mathcal{M}_1^{lin}&(\brE_1)\label{SystemEqOrdre3Degre3E0nulE1} \\
&+\omega_I^2\Big(2\dfl\brE_{-2},\brE_3\ffl+2\dfl\brE_{-1},\brE_2\ffl\nonumber \\
&+6\dfl\brE_{-3},\brE_1,\brE_3\ffl+3\dfl\brE_{-1},\brE_{-1},\brE_3\ffl+3\dfl\brE_{-3},\brE_2,\brE_2\ffl\nonumber \\
&+6\dfl\brE_{-2},\brE_1,\brE_2\ffl+3\dfl\brE_{-1},\brE_1,\brE_1\ffl\Big)=\frac{-i\,\omega_I}{\eps_0}\brJ_1,\nonumber \\
\nonumber \\
\mathcal{M}_2^{lin}&(\brE_2)\label{SystemEqOrdre3Degre3E0nulE2} \\
&+(2\omega_I)^2\Big(2\dfl\brE_{-1},\brE_3\ffl+\dfl\brE_1,\brE_1\ffl\nonumber \\
&+6\dfl\brE_{-3},\brE_2,\brE_3\ffl+6\dfl\brE_{-2},\brE_1,\brE_3\ffl+3\dfl\brE_{-2},\brE_2,\brE_2\ffl\nonumber \\
&+6\dfl\brE_{-1},\brE_1,\brE_1\ffl\Big)=0,\nonumber \\
\nonumber \\
\mathcal{M}_3^{lin}&(\brE_3)\label{SystemEqOrdre3Degre3E0nulE3} \\
&+(3\omega_I)^2\Big(2\dfl\brE_1,\brE_2\ffl\nonumber \\
&+3\dfl\brE_{-3},\brE_3,\brE_3\ffl+6\dfl\brE_{-2},\brE_2,\brE_3\ffl+6\dfl\brE_{-1},\brE_1,\brE_3\ffl\nonumber \\
&+3\dfl\brE_{-1},\brE_2,\brE_2\ffl+\dfl\brE_1,\brE_1,\brE_1\ffl\Big)=0.\nonumber
\end{align}
\end{subequations}
The optical effects that this systems describes are the same than
the ones present in the $d_1=3$ degree: optical Kerr-effect,
depletion of the pump beam, second and third harmonic generations.

Once again, only energy considerations, presented in the following
section, can justify that we keep terms like
$\dfl\brE_2,\brE_2,\brE_{-3}\ffl$, that are, of course, several
orders of magnitude smaller than $\dfl\brE_1,\brE_1,\brE_{-1}\ffl$
(see the equation (\ref{SystemEqOrdre3Degre3E0nulE1})).

\vspace{.2cm}

This last example closes this section. The introduction of the
degree allows to present systems whose solutions can be numerically
investigated. Moreover, we see that, from the physical point of view
(i.e., with the degree $d_1$), the optical Kerr effect appears
simultaneously to the depletion of the pump beam or the third
harmonic generation. This kind of statements is exactly the one we
looked for; this is why we consider that, when restricting to
harmonic processes with only one angular frequency in the incident
field, we have fulfil our general aim. In \cite{bGodard10} is
exposed a method to generalize this when the scattered field
oscillate at a nonharmonic frequencies or when the source oscillate
at two frequencies. But this presents several drawbacks: first, this
does not allow to treat phenomena described by a continuous set of
frequencies, which are of course the most interesting ones from the
physical point of view, and secondly, the systems obtained quickly
become huge; solving them requires new hypotheses, so that it is
like going back to the starting point: we have to impose constraints
that depend on the particular situation we want to study.

\begin{remONL}{The optical Kerr effect}\label{RemOpticalKerrEffect}

We said that the $n$-th harmonic generation is better studied in the
$n$-th order of nonlinearity and the $n$-th degree; nevertheless,
the third order of nonlinearity in the degree $d_{\infty}=1$ is also
interesting. It consists of the single equation

\begin{equation}\label{SystemEqOrdre3Degre1E0nul}
\mathcal{M}_1^{lin}(\brE_1)+3\omega_I^2\dfl\brE_{-1},\brE_1,\brE_1\ffl=\frac{-i\,\omega_I}{\eps_0}\brJ_1.
\end{equation}
This is the optical Kerr effect. We repeat that we can argue that
this system discards the term $\dfl\brE_1,\brE_1,\brE_1\ffl$ while
keeping $\dfl\brE_{-1},\brE_1,\brE_1\ffl$, thus `neglecting' \emph{a
priori} that $\|\brE_{-1}\|=\|\brE_1\|$. Nevertheless, in some
cases, we can be concerned only to what happens in the fundamental
frequency, and so we disregard the harmonic generation. This
naturally leads to the study of this optical Kerr effect. We note
that, dealing with this equation, we neglect the counter-reaction of
the harmonics on the fundamental component, i.e., we neglect the
term:

$$2\dfl\brE_{-2},\brE_3\ffl+2\dfl\brE_{-1},\brE_2\ffl+6\dfl\brE_{-3},\brE_1,\brE_3\ffl$$
$$+3\dfl\brE_{-1},\brE_{-1},\brE_3\ffl+3\dfl\brE_{-3},\brE_2,\brE_2\ffl
+6\dfl\brE_{-2},\brE_1,\brE_2\ffl$$%
that contributes to $\hat{\brP}(\omega_I)$. For non-centrosymmetric
media, in which the second order polarization vector fields vanish,
this approximation can be dangerous, since it is not always clear
that

$$2\|\dfl\brE_{-1},\brE_2\ffl\|\ll3\|\dfl\brE_{-1},\brE_1,\brE_1\ffl\|.$$
However, for centrosymmetric media, the term
$\|\dfl\brE_{-1},\brE_1,\brE_1\ffl\|$ must be the leading one among
the terms that contributes to $\hat{\brP}^{(3)}(\omega_I)$; this
justifies that the $d_{\infty}=1$ system is an approximation of the
$d_{\infty}=3$ system. Hence, the way we obtained this equation
shows that the optical Kerr effect appears naturally, and that it
does make sense to study a \emph{harmonic} nonlinear effect.

We will go back, in the next section, to that simple system as a
basic example of energy conservation.
\end{remONL}

\section{Electric energy considerations in nonlinear
media}\label{SectionEnergy}

\subsection{The stochastic mean of the electric power density}

Before dealing with energy conservation in nonlinear optics, we have
to give a rigorous definition of what we mean by energy transfer.
Indeed, the components oscillating at the frequency of the incident
field generate the other components, and so will loose some energy.
The question is thus the following: when this harmonic generation
occur with no loss of electric energy, that is, without Joule
effect. Stated in an other way: when does the energy in the electric
sources equal the energy in the scattered field?

To this aim, we go back to the definition of the energy-momentum
quadrivector, or, more precisely, its density, denoted by
$(c\,\mathcal{I},W)$. By definition, it is divergence-free, so that
in a Cartesian coordinate frame, where

$$\nabla^4\cdot:=(-c\partial_x\,\,-c\partial_y\,\,-c\partial_z\,\,\partial_t)\cdot,$$
we have

\begin{equation*}%\label{ConsImEn}
-c\,\nabla\cdot(c\,\mathcal{I})+\partial_tW=0.
\end{equation*}
From now on, we consider only the electromagnetic part
$(c\,\mathcal{I}_{em},W_{em})$ of $(c\,\mathcal{I},W)$. We thus say
that no transfer between electromagnetic energy-momentum and other
form of energy-momentum occurs if

\begin{equation*}%\label{ConsImEnEM}
-c\,\nabla\cdot(c\,\mathcal{I}_{em})+\partial_tW_{em}=0.
\end{equation*}
For geometric reasons, and for the respect of the units of each
field, the Poynting vector field, $\mathcal{P}:=\brE\times\brH$, is
defined to be the density of electromagnetic momentum
$c\,\mathcal{I}_{em}$ multiplied by the factor $-c$. The
electromagnetic energy-momentum conservation is thus

$$\nabla\cdot\mathcal{P}+\partial_tW_{em}=0.$$
This relation allows to define the electromagnetic energy density
$W_{em}$. Indeed, Maxwell's equations give:

\begin{align*}
\partial_tW_{em}&=-\nabla\cdot\mathcal{P}\\
&=(\partial_t\brB)\cdot\brH+(\partial_t\brD)\cdot\brE+\brE\cdot\brJ.
\end{align*}
We define the electric energy density $W_e$ by

$$\partial_tW_e=(\partial_t\brD)\cdot\brE.$$

We write the electric vector field evaluated at the point $\brs$ and
the time $t$ as (we need an absolute convergence of this series:
indeed, in this case, the series converge uniformly with respect to
the time variable and thus we will be allowed to change the integral
operation with the sum operation)

\begin{equation*}
\brE(\brs,t)=\sum_{p\in\Z}\brE_p(\brs)e^{-ip\omega_It}.
\end{equation*}
Once again, we will give all the details in the second order of
nonlinearity, and then present more general formulae. So let us
write

\begin{align*}
\brD(t)&=\eps_0\brE(t)+\brP(t) \\
&=\eps_0\sum_{p\in\Z}\Big(\brE_p+\dfl\brE_p\ffl+\sum_{q\in\Z}\dfl\brE_q,\brE_{p-q}\ffl\Big)e^{-ip\omega_It}.
\end{align*}
The electric energy density is therefore\footnote{Up to now, the dot
product was on $\R^3$. We extend it on $\C^3$ since we decompose the
vectors in their Fourier components. We choose to take it linear in
both variables, and so it will be symmetric - this will not bring
problems since the nondegeneracy will never be exploited.}

\begin{align*}
W_e(t)&=W_e(0) \\
&\quad+\eps_0\sum_{(p,q)\in\Z^*\times\Z}\frac{q}{p}\brE_{p-q}\cdot\big(\brE_q+\dfl\brE_q\ffl+\sum_{r\in\Z}\dfl\brE_r,\brE_{q-r}\ffl\big)(e^{-ip\omega_It}-1) \\
&\quad+\eps_0\sum_{p\in\Z}-ip\omega_I\brE_{-p}\cdot\big(\brE_p+\dfl\brE_p\ffl+\sum_{q\in\Z}\dfl\brE_q,\brE_{p-q}\ffl\big)t.
\end{align*}
The first term in this expression is a constant without any physical
significance, the second one is oscillating and the last one
diverges linearly with $t$. For large $t$ (large with respect to the
period associated to the fundamental angular frequency $\omega_I$),
only the last term is relevant. This leads to define the following
quantity, which is the stochastic mean of the electric power
density:

\begin{align*}%\label{eq<partial_tW_e>}
<\partial_tW_e>:&=\lim_{T\rightarrow\infty}\frac{1}{T}\int_0^Tdt\,\partial_tW_e\nonumber \\
&=\eps_0\sum_{p\in\Z}\Big(-ip\omega_I\brE_{-p}\cdot\big(\brE_p+\dfl\brE_p\ffl+\sum_{q\in\Z}\dfl\brE_q,\brE_{p-q}\ffl\big)\Big).
\end{align*}
The term
$\displaystyle{\eps_0\big(\brE_p+\dfl\brE_p\ffl+\sum_{q\in\Z}\dfl\brE_q,\brE_{p-q}\ffl\big)}$,
is the expansion of $\brD_p$ as
$\eps_0\brE_p+\brP^{(1)}_p+\brP^{(2)}_p$. We recall that, by
construction, the answer of the medium is described by the
polarization vector field $\brP$; this is coherent with the fact
that the sum
$\displaystyle{\sum_{p\in\Z}-ip\omega_I\brE_{-p}\cdot\brE_p}$ (where
no susceptibility tensor appears) vanishes - we map $p\mapsto -p$
and use the Hermitian symmetry of the duality product on $\C^3$. We
thus have shown that, in vacuum, the stochastic mean of the electric
power density
$\displaystyle{\eps_0\sum_{p\in\Z}-ip\omega_I\brE_{-p}\cdot\brE_p}$
vanishes.

Going back to an arbitrary order of nonlinearity, we have

\begin{align}\label{Eqpartial_tWe}
<\partial_tW_e>&=<(\partial_t\brP)\cdot\brE> \\
&=\eps_0\sum_{p\in\Z}\Big(-ip\omega_I\brE_{-p}\cdot\big(\dfl\brE_p\ffl+\sum_{q\in\Z}\dfl\brE_q,\brE_{p-q}\ffl+\cdots\big)\Big)\nonumber \\
&=\sum_{n\in\N}\sum_{p\in\Z}-ip\omega_I\brE_{-p}\cdot\brP_p^{(n)}\nonumber \\
&=\sum_{p\in\Z}-ip\omega_I\brE_{-p}\cdot\brP_p.\nonumber
\end{align}
We remark that this expression is a straightforward generalization
of the stochastic mean of the electric power density in the linear
regime. It is also clear that this expression is real - taking the
complex conjugate is equivalent to map $p$ to $-p$ (and $q$ to $-q$
in the second line), and thus does not change the sum.

\subsection{A symmetry of the susceptibility tensors and lossless media}

We did not found a relation between the susceptibility tensors that
exactly translate the property of a medium to be lossless.
Nevertheless, we found one implication: if a certain symmetry
between the susceptibility tensors is satisfied then the medium is
lossless. This subsection is devoted to the presentation of this
criterion.

The first hypothesis that we will do is to consider that the
electromagnetic power density vanishes in each order (we won't
repeat `the stochastic mean of' each time, but it is always
understood). We thus define

$$<\partial_tW_e^{(n)}>:=\sum_{p\in\Z}-ip\omega_I\brE_{-p}\cdot\brP_p^{(n)}.$$
The third equality in (\ref{Eqpartial_tWe}), shows that

$$<\partial_tW_e>=\sum_{n\in\N}<\partial_tW_e^{(n)}>.$$

Using the Hermitian symmetry of the Fourier component of the
electric vector field and of the polarization vector field, it is
straightforward to obtain the expression

\begin{equation}\label{Eqpartial_tWenIM}
<\partial_tW_e^{(n)}>=2\omega_I\IM\{\sum_{p\in\N}p\brE_{-p}\cdot\brP^{(n)}_p\}.
\end{equation}

\subsubsection{In the first order}

In the first order, we have

\begin{align*}
<\partial_tW^{(1)}_e>:&=\eps_0\sum_{p\in\Z}-ip\omega_I\brE_{-p}\cdot\dfl\brE_p\ffl \\
&=-i\eps_0\omega_I\sum_{p\in\N}p\brE_{-p}\cdot\Big(\chi^{(1)}(p\omega_I)-\chi^{(1)\,T}(-p\omega_I)\Big)\brE_p.
\end{align*}
Here we make the second assumption: we assume that each term of this
sum vanishes. This leads to the conclusion that no electromagnetical
energy is lost in the first order if the susceptibility tensor field
(on $\R^3\times\R$) $\chi^{(1)}$ is Hermitian, i.e.:

\begin{equation}\label{CritEn1}
\chi^{(1)}(p\omega_I)=\chi^{(1)\,T}(-p\omega_I).
\end{equation}

The conclusion is not exactly the well-known fact from undergraduate
lessons: if the medium is linear - so that the fields oscillate at
only one angular frequency $\omega_I$ -, it is lossless from the
electrical point of view if and only if the susceptibility tensor
field (on $\R^3$) $\chi^{(1)}(\cdot,\omega_I)$ is Hermitian. Here,
the medium is nonlinear - so that the fields oscillate at every
angular frequency $p\omega_I$ for $p$ in $\Z^*$ -, and we showed
that is lossless from the electrical point of view in the first
order if the susceptibility tensor field $\chi^{(1)}$ is Hermitian.

\subsubsection{In the second order}

\paragraph{The general criterion}

We now go to the second order:

\begin{align*}
<\partial_tW^{(2)}_e>&=\sum_{p\in\Z}-ip\omega_I\brE_{-p}\cdot\brP_p^{(2)} \\
&=\eps_0\sum_{(p,q)\in\Z^2}-ip\omega_I\brE_{-p}\cdot\dfl\brE_q,\brE_{p-q}\ffl.
\end{align*}

Some lines of computation show that
$$<\partial_tW^{(2)}_e>=
\omega_I\eps_0\IM\{\sum_{\substack{q\in\N \\ 0\leqslant r\leqslant
q}}d_2(q,r)d_2(r,0)\brE_{-(q+r)}\cdot\xi^{(2)}_{q,r}\brE_r\brE_q\},$$
with

\begin{align*}
\xi^{(2)}_{q,r}&:=r\big(\chi^{(2)}(r\omega_I,q\omega_I)-\chi^{(2)\,T_{12}}(-(q+r)\omega_I,q\omega_I)\big) \\
&+q\big(\chi^{(2)}(r\omega_I,q\omega_I)-\chi^{(2)\,T_{13}}(r\omega_I,-(q+r)\omega_I)\big).
\end{align*}
where the partial transposition notation is defined by
$\brv_a\cdot\chi^{(2)\,T_{12}}\brv_b\brv_c:=\brv_b\cdot\chi^{(2)}\brv_a\brv_c$
for all triples $(\brv_a,\brv_b,\brv_c)$ of vectors in $\R^3$; with
indices, this gives
$(\chi^{(2)\,T_{12}})^i_{\,\,\,i_1i_2}=\chi^{(2)\,\,\,\,\,i}_{\quad
i_1\,\,\,i_2}$. In the same way we write
$\brv_a\cdot\chi^{(2)\,T_{13}}\brv_b\brv_c:=\brv_c\cdot\chi^{(2)}\brv_b\brv_a$
and so
$(\chi^{(2)\,T_{13}})^i_{\,\,\,i_1i_2}=\chi^{(2)\quad\;i}_{\quad
i_2i_1}$. Also, $d_2(\cdot,\cdot)$ is the degeneracy function of a
set of two elements:

\begin{equation*}
  d_2(a,b)=
     \begin{cases}
        2, & a\neq b \\
        1, & a=b.
     \end{cases}
\end{equation*}

For $<\partial_tW^{(2)}_e>$ to vanish, it is thus sufficient that
each term of the sum vanishes (this is similar to what we did in the
first order). We now introduce the third assumption: the factor of
$r$ and the one of $q$ vanish independently. Thus, no transfer of
electric energy to an other form of energy is possible if the tensor
field $\chi^{(2)}$ satisfies, for any $(q,r)$ in $\Z^2$ such that
$q>0$ and $0\leqslant r\leqslant q$, the relations:

\begin{equation*}
     \begin{cases}
        \chi^{(2)}(r\omega_I,q\omega_I)=\chi^{(2)\,T_{12}}(-(q+r)\omega_I,q\omega_I) \\
        \chi^{(2)}(r\omega_I,q\omega_I)=\chi^{(2)\,T_{13}}(r\omega_I,-(q+r)\omega_I).
     \end{cases}
\end{equation*}

%\begin{table}[htbp]
%   \begin{center}
%\begin{tabular}{|c|}
%  \hline
%  \\
%  $\chi^{(2)}(r\omega_I,q\omega_I)=\chi^{(2)\,T_{12}}(-(q+r)\omega_I,q\omega_I)$ \\
%  \\
%  et \\
%  \\
%  $\chi^{(2)}(r\omega_I,q\omega_I)=\chi^{(2)\,T_{13}}(r\omega_I,-(q+r)\omega_I)$ \\
%  \\
%  \hline
%\end{tabular}
%  \caption{Conditions sur le tenseur $\chi^{(2)}$ qui garantissent l'absence de pertes par effet Joule à l'ordre deux.
%\label{TablePasDePertesChi2}}
%   \end{center}
%\end{table}

A more symmetric way of writing these conditions appear if we define
$p:=-(q+r)$ and we write explicitly, as an argument of the
susceptibility tensor, the angular frequency for which $\chi^{(2)}$
contribute in $\brP^{(2)}$. Thus, from now on, any of the following
expression has the same meaning:
$\chi^{(n)}(\brs;\omega_0;\omega_1,\cdots,\omega_n)$,
$\chi^{(n)}(\brs,\omega_1,\cdots,\omega_n)$,
$\chi^{(n)}(\omega_0;\omega_1,\cdots,\omega_n)$,
$\chi^{(n)}(\omega_1,\cdots,\omega_n)$, and when $\chi^{(n)}$ is
evaluated on $n+1$ angular frequencies $\omega_0$, $\cdots$,
$\omega_n$, it is understood that
$\omega_0=\omega_1+\cdots+\omega_n$. With this notation, a medium is
lossless from the electrical point of view in the second order if

\begin{equation}\label{CritEn2}
\chi^{(2)}(-p\omega_I;r\omega_I,q\omega_I)=\chi^{(2)\;T_{12}}(-r\omega_I;p\omega_I,q\omega_I)=\chi^{(2)\;T_{13}}(-q\omega_I;r\omega_I,p\omega_I).
\end{equation}
So no electric energy loss is guaranteed if the symmetry that
consists in permuting the indices together with the frequency
variables holds for the $\chi^{(n)}$. This result appears in the
book \cite{Popov}, but these authors do not consider the cascade
effects that appear in harmonic generations. Moreover, the
assumptions needed to derive these results are not explicitly given.

We note that (contrary to what is affirmed in \cite{Boyd}, p.34) we
can have a lossless medium with complex-valued susceptibility tensor
fields. This has been checked numerically by the authors, with
simulations similar to the ones given in \cite{aGodard11b}.

\paragraph{Kleinman's relations}

If the medium is instantaneous, then the response function $R^{(n)}$
is of the kind
$\displaystyle{A\bigotimes_{i\in\{1,\cdots,n\}}\delta(t_i)}$ where
$A$ is a function that depends only on the space variables. Then the
susceptibility tensor fields do not depend on the angular
frequencies:

$$\chi^{(n)}_{inst}:=\chi^{(n)}(p_1\omega_I,\cdots,p_n\omega_I)\qquad\forall (p_1,\cdots,p_n)\in\Z^n.$$
Thus, in the first order, the polarization is

\begin{align*}
\brP^{(1)}(t)&=\eps_0\int_{-\infty}^{\infty}d\omega_1\chi^{(1)}_{inst}\hat{\brE}(\omega_1)e^{-i\omega_1t} \\
&=\eps_0\chi^{(1)}_{inst}\int_{-\infty}^{\infty}d\omega_1\hat{\brE}(\omega_1)e^{-i\omega_1t} \\
&=\eps_0\chi^{(1)}_{inst}\brE(t).
\end{align*}
Since $\chi^{(1)}_{inst}$, when contracted with a real vector, gives
a real vector, it is real.

In the same way, in the second order, we have

\begin{align*}
\brP^{(2)}(t)&=\eps_0\int_{-\infty}^{\infty}d\omega_1\int_{-\infty}^{\infty}d\omega_2\,\chi^{(2)}_{inst}
\hat{\brE}(\omega_1)\hat{\brE}(\omega_2)e^{-i(\omega_1+\omega_2)t} \\
&=\eps_0\chi^{(2)}_{inst}\int_{-\infty}^{\infty}d\omega_1\,\hat{\brE}(\omega_1)e^{-i\omega_1t}\int_{-\infty}^{\infty}d\omega_2\,\hat{\brE}(\omega_2)e^{-i\omega_2t} \\
&=\eps_0\chi^{(2)}_{inst}\brE(t)\brE(t)
\end{align*}
and $\chi^{(2)}_{inst}$ is also real.

Reporting the criteria (\ref{CritEn1}) and (\ref{CritEn2}) in this
context, we deduce that an instantaneous nonlinear medium is
lossless if the susceptibility tensor fields that characterize it
satisfy

\begin{equation*}
     \begin{cases}
        \textrm{in the first order} : & \chi^{(1)}_{inst}=\chi^{(1)\;T}_{inst} \\
        \textrm{in the second order} : & \chi^{(2)}_{inst}=\chi^{(2)\;T_{12}}_{inst}=\chi^{(2)\;T_{13}}_{inst}.
     \end{cases}
\end{equation*}
Because the transposition $(12)$ generate the symmetric group
$\mathcal{S}_2$, the transpositions $(12)$ and $(13)$ generate
$\mathcal{S}_3$, etc., where $\mathcal{S}_{n+1}$ acts as permutation
of the components of $\chi^{(n)}$, we conclude that completely
symmetric and frequency independent susceptibility tensor
characterize lossless media. This is the famous result of
D.A.Kleinman (\cite{Kleinman}).

\paragraph{In the lower degrees}

In the second order $1\leqslant n\leqslant2$ and the degree
$d_{\infty}=2$, we have

\begin{align*}
\brP^{(1)}_1&=\eps_0\dfl\brE_1\ffl, \\
\brP^{(1)}_2&=\eps_0\dfl\brE_2\ffl, \\
\brP^{(2)}_1&=2\eps_0\dfl\brE_{-1},\brE_2\ffl, \\
\brP^{(2)}_2&=\eps_0\dfl\brE_1,\brE_1\ffl.
\end{align*}
Hence, an explicit expansion of (\ref{Eqpartial_tWenIM}) gives

$$<\partial_tW_e>=2\eps_0\omega_I\IM\{\brE_{-1}\cdot(\dfl\brE_1\ffl+2\dfl\brE_{-1},\brE_2\ffl)+2\brE_{-2}\cdot(\dfl\brE_2\ffl+\dfl\brE_1,\brE_1\ffl)\}.$$
We first treat the first order of nonlinearity. So let us assume
that $\chi^{(1)}(\brs,\omega_I)$ is Hermitian for all $\brs$ in
$\R^3$, in the sense that its transpose conjugate is equal to
itself; we then have

\begin{align*}
\brE_{-1}\cdot\dfl\brE_1\ffl:&=\brE_{-1}\cdot\chi^{(1)}(\omega_I)\brE_1 \\
&=\brE_1\cdot\chi^{(1)\,T}(\omega_I)\brE_{-1} \\
&=\overline{\brE_{-1}}\cdot\overline{\chi^{(1)}(\omega_I)}\overline{\brE_1} \\
&=\overline{\brE_{-1}\cdot\chi^{(1)}(\omega_I)\brE_1} \\
&=:\overline{\brE_{-1}\cdot\dfl\brE_1\ffl},
\end{align*}
We deduce from this that $\IM\{\brE_{-1}\cdot\dfl\brE_1\ffl\}$
vanishes: this is the condition (\ref{CritEn1}) on electrical energy
conservation in the first order. A similar computation shows that if
$\chi^{(1)}(\brs,2\omega_I)$ is Hermitian for all $\brs$ in $\R^3$,
then $\IM\{\brE_{-2}\cdot\dfl\brE_2\ffl\}$ vanishes.

Considering now the second order of nonlinearity, we suppose that
the second order susceptibility tensor field satisfies

\begin{equation}\label{CriterionEnOrder2degree2}
\chi^{(2)}(-p\omega_I;r\omega_I,q\omega_I)=\chi^{(2)\,T_{13}}(-q\omega_I;r\omega_I,p\omega_I)
\end{equation}%
for all $(q,r)$ in $\Z^2$ - we will use only
$\chi^{(2)\,T_{13}}(\omega_I;-\omega_I,2\omega_I)=\chi^{(2)}(-2\omega_I;-\omega_I,-\omega_I)$.
Then we have

\begin{align*}
<\partial_tW_e^{(2)}>&=4\eps_0\omega_I\IM\{\brE_{-1}\cdot\dfl\brE_{-1},\brE_2\ffl+\brE_{-2}\cdot\dfl\brE_1,\brE_1\ffl\} \\
&=4\eps_0\omega_I\IM\{\brE_{-1}\cdot\chi^{(2)}(-\omega_I,2\omega_I)\brE_{-1}\brE_2+\brE_{-2}\cdot\chi^{(2)}(\omega_I,\omega_I)\brE_1\brE_1\} \\
&=4\eps_0\omega_I\IM\{\brE_2\cdot\chi^{(2)\,T_{13}}(-\omega_I,2\omega_I)\brE_{-1}\brE_{-1}+\brE_{-2}\cdot\chi^{(2)}(\omega_I,\omega_I)\brE_1\brE_1\} \\
&=4\eps_0\omega_I\IM\{\brE_2\cdot\chi^{(2)}(-\omega_I,-\omega_I)\brE_{-1}\brE_{-1}+\brE_{-2}\cdot\chi^{(2)}(\omega_I,\omega_I)\brE_1\brE_1\} \\
&=8\eps_0\omega_I\IM\{\RE\{\brE_2\cdot\chi^{(2)}(-\omega_I,-\omega_I)\brE_{-1}\brE_{-1}\}\} \\
&=0,
\end{align*}
where the last but one line is obtained by the Hermitian symmetry of
the electric vector field and of the susceptibility tensor fields.
We thus see that the criterion (\ref{CriterionEnOrder2degree2}) is a
sufficient condition for the stochastic mean of the electrical
energy density at the second order to vanish.

This is why we argued, in the subsection \ref{SubsecNLorder2}, that
though the degree $d_1$ is the most satisfying one from the physical
point of view, even if it leads to consider $\brE_1$ as a tank, the
degree $d_{\infty}$ allows to consider lossless media. This
advantage can be a good test to check the coherence of numerical
simulations.

Lastly, we note that (\ref{CriterionEnOrder2degree2}) is weaker than
(\ref{CritEn2}); this is due to the degeneracy in the angular
frequencies in the order two and degree two. The reader can check
that the full criterion (\ref{CritEn2}) is necessary if we consider
the order two in the degree three.

\subsubsection{In the third order}

\paragraph{The general criterion}

In the third order, we have

$$<\partial_tW^{(3)}_e>:=\eps_0\sum_{(q,r,s)\in\Z^3}-iq\omega_I\brE_{-q}\cdot\dfl\brE_r,\brE_s,\brE_{q-r-s}\ffl.$$
We can show (\cite{bGodard10}) that if the following conditions are
satisfied

\begin{equation*}
     \begin{cases}
        \chi^{(3)}(-p\omega_I;q\omega_I,r\omega_I,s\omega_I)=\chi^{(3)\;T_{12}}(-q\omega_I;p\omega_I,r\omega_I,s\omega_I) \\
        \chi^{(3)}(-p\omega_I;q\omega_I,r\omega_I,s\omega_I)=\chi^{(3)\;T_{13}}(-r\omega_I;q\omega_I,p\omega_I,s\omega_I) \\
        \chi^{(3)}(-p\omega_I;q\omega_I,r\omega_I,s\omega_I)=\chi^{(3)\;T_{14}}(-s\omega_I;q\omega_I,r\omega_I,p\omega_I)
     \end{cases}
\end{equation*}
then a medium is lossless from the electrical point of view in the
third order.

%\begin{table}[htbp]
%   \begin{center}
%\begin{tabular}{|c|}
%  \hline
%  \\
%  $\chi^{(3)}(-p\omega_I;q\omega_I,r\omega_I,s\omega_I)=\chi^{(3)\;T_{12}}(-q\omega_I;p\omega_I,r\omega_I,s\omega_I)$ \\
%  $=\chi^{(3)\;T_{13}}(-r\omega_I;q\omega_I,p\omega_I,s\omega_I)=\chi^{(3)\;T_{14}}(-s\omega_I;q\omega_I,r\omega_I,p\omega_I)$ \\
%  \\
%  \hline
%\end{tabular}
%  \caption{Conditions sur le tenseur $\chi^{(3)}$ qui garantissent l'absence de pertes par effet joule à l'ordre trois.
%\label{TablePasDePertesChi3}}
%   \end{center}
%\end{table}

\paragraph{In the lower degrees}

As exposed in the remark \ref{RemOpticalKerrEffect}, the system of
propagation equations (\ref{EqProgEn3}) in the degree $d_{\infty}=1$
reduces to the single equation that describes the optical Kerr
effect:

\begin{equation}\tag{\eqref{SystemEqOrdre3Degre1E0nul}}
\mathcal{M}_1^{lin}(\brE_1)+3\omega_I^2\dfl\brE_{-1},\brE_1,\brE_1\ffl=\frac{-i\,\omega_I}{\eps_0}\brJ_1.
\end{equation}

Since

$$\brP^{(3)}_1=\eps_0\dfl\brE_{-1},\brE_1,\brE_1\ffl,$$
the formula (\ref{Eqpartial_tWenIM}) gives

\begin{align*}
<\partial_tW^{(3)}_e>_{\textrm{Kerr}}&=2\omega_I\IM\{\sum_{p\in\N}p\brE_{-p}\cdot\brP^{(3)}_p\} \\
&=6\eps_0\omega_I\IM\{\brE_{-1}\cdot\dfl\brE_{-1},\brE_1,\brE_1\ffl\}
\end{align*}

%%If, furthermore, the medium is isotropic (see \cite{Boyd} for a
%%classical reference, or \cite{bGodard10} for more definition), then
%%we have
%%
%%\begin{align*}
%%<\partial_tW^{(3)}_e>_{\textrm{K\,opt}}&=-3\eps_0\omega_I\IM\{\brE_1 \\
%%&\quad\cdot\big(\chi^{(3)}(\omega_I,-\omega_I,-\omega_I)-\chi^{(3)\;T_{13}}(\omega_I,\omega_I,-\omega_I)\big)\brE_1\brE_{-1}\brE_{-1}\}
%%\end{align*}

Assume furthermore that the medium is invariant along one axis (say
$(O,z)$), and that if a beam linearly polarized along $(O,z)$
impinges on this medium, then the scattered light is also polarized
along $(O,z)$. Then, the only components of the susceptibility
tensors that matter are $\chi^{(1)\,z}_{\qquad z}$ and
$\chi^{(3)\,z}_{\qquad zzz}$. Let us denote $\brE_1(x,y,z)$ by
$u_1(x,y)\hat{z}$.

%On remarque que, dans ces conditions, les transpositions sur
%$\chi^{(n)}$ sont triviales, (ce qui implique, par la symétrie de
%permutation, que l'on peut échanger l'ordre des arguments
%fréquentiels à notre guise) et l'on peut récrire la condition
%précédente sous la
%forme :%\footnote{En utilisant la définition de la transposée et la
%%symétrie intrinsèque, on a
%%\hbox{$\brE_1\chi^{(3)\;T_{13}}(\omega_I,\omega_I,-\omega_I)\brE_1\brE_{-1}\brE_{-1}=\brE_{-1}\chi^{(3)}(\omega_I,\omega_I,-\omega_I)$}
%%=\hbox{$\brE_{-1}\chi^{(3)}(\omega_I,-\omega_I,\omega_I)\brE_1\brE_{-1}\brE_1=\brE_1\chi^{(3)\;T_{14}}(\omega_I,-\omega_I,\omega_I)$}.}

In this case, we have

\begin{align*}
<\partial_tW^{(3)}_e>_{\textrm{Kerr}}&=6\eps_0\omega_I\IM\{u_{-1}\chi^{(3)\,z}_{\qquad
zzz}(-\omega_I,\omega_I,\omega_I)u_{-1}u_1u_1\} \\
&=6\eps_0\omega_I\IM\{\chi^{(3)\,z}_{\qquad
zzz}(-\omega_I,\omega_I,\omega_I)\}|u_1|^4.
\end{align*}

In other words, with all these assumption, there are no electrical
energy loss in the third order \emph{if and only if} the
susceptibility tensor $\chi^{(3)}$ is real-valued.

\subsection{In the $n$-th order}

The general proof that a sufficient condition for a medium to be
lossless in the $n$-th order can be found in \cite{bGodard10}. The
result is that

\begin{align*}
&<\partial_tW_e^{(n)}>=\sum_{p\in\Z}-ip\omega_I\brE_{-p}\cdot\brP_p^{(n)} \\
&\quad=-i\eps_0\omega_I\sum_{(p_1,\cdots,p_n)\in\Z^n}(p_1+\cdots+p_n)\brE_{-(p_1+\cdots+p_n)}\cdot\dfl\brE_{p_1},\cdots,\brE_{p_n}\ffl.
\end{align*}
vanishes if

$$\chi^{(n)\;T_{1\,j+1}}(-p_j\omega_I;p_0\omega_I,\cdots,p_{j-1}\omega_I,p_{j+1}\omega_I,\cdots,p_n\omega_I)=$$
$$\chi^{(n)}(-p_0\omega_I;p_1\omega_I,\cdots,p_n\omega_I)\qquad\forall j\in\{1,\cdots,n\}$$
where $p_0:=-(p_1+\cdots+p_n)$. We can check that the criteria given
for the lowest order fulfils this set of conditions.

Once again, we repeat that we did not find a necessary and
sufficient criterion: we supposed that each order vanishes
independently, that the term $<\partial_t(W_e^{(n)}>$ vanishes for
each set $(-p_0,p_1,\cdots,p_n)$, with $p_0=-(p_1+\cdots+p_n)$,
where $\brE_{p_0}$, $\brE_{p_1}$, $\cdots$, $\brE_{p_n}$ appear in
$\brE_{-p}\cdot\brP^{(n)}_p$, and finally that, within each such
set, the factors of $p_1$, etc., $p_n$ vanish separately. These
factors are of course the terms

$$\chi^{(n)\;T_{1\,j+1}}(-p_j\omega_I;p_0\omega_I,\cdots,p_{j-1}\omega_I,p_{j+1}\omega_I,\cdots,p_n\omega_I)$$
$$-\chi^{(n)}(-p_0\omega_I;p_1\omega_I,\cdots,p_n\omega_I),$$
justifying the general criterion.

This general result appears in \cite{Popov}; we hope that the
physical meaning of these equations, as well as the assumptions we
were led to carry out, have been clarified.

%\begin{table}[htbp]
%   \begin{center}
%\begin{tabular}{|c|}
%  \hline
%  \\
%  $\chi^{(n)\;T_{1\,j+1}}(-p_j\omega_I;p_0\omega_I,\cdots,p_{j-1}\omega_I,p_{j+1}\omega_I,\cdots,p_n\omega_I)$ \vspace{.1cm}\\
%  $=$ \\
%  $\chi^{(n)}(-p_0\omega_I;p_1\omega_I,\cdots,p_n\omega_I)$ \vspace{.1cm}\\
%  $\qquad\forall j\in\{1,\cdots,n\}$ \\
%  \\
%  \hline
%\end{tabular}
%  \caption{Conditions sur le tenseur $\chi^{(n)}$ qui garantissent l'absence de pertes par effet joule à l'ordre $n$.
%\label{TablePasDePertesChin}}
%   \end{center}
%\end{table}

%On peut également signaler que R.W. Boyd "montre", aux pages 34-36
%de \cite{Boyd}, avec des arguments qui nous semblent insuffisants,
%que si un milieu est sans perte, alors il satisfait au critère
%énergétique (il faut également utiliser la symétrie de permutation
%intrinsèque, valide pour tous les milieux). Nous remarquons que,
%d'une part, ce que nous avons montré est l'implication inverse, et
%que d'autre part, ce résultat nous paraît incompatible avec sa
%conception d'un milieu sans perte, qui est selon lui un milieu où
%les tenseurs sont à image réelle. Nous verrons dans le chapitre
%\ref{ChapitreHarmGenSimul} des exemples où la susceptibilité est
%décrite par des nombres complexes et où, pour autant, il y a
%conservation de l'énergie électrique.

\section{Conclusion and outlook}

This paper explores a new route to obtain the propagation equation
systems governing nonlinear interactions of light in matter. The aim
is to obtain equations that do not depend on the particular effect,
the particular involved material or the particular situation we want
to study. This aim is partly fulfilled, especially when treating
harmonic generation. For this, the introduction of the degree is of
major importance. To our knowledge, no similar notion has been
introduced in the literature.

The paper \cite{aGodard11b} is devoted to numerical results; the
system (\ref{SystemEqOrdre3Degre3E0nul}) is chosen for its
interesting effects (second and third harmonic generation, optical
Kerr effect, depletion of the pump beam) and because it allows to
make a power balance. Of course, only an experimental test could
invalidate the theoretical as well as the numerical models. Though
these experiments should not be difficult to carry out, we have to
confess, without being to much ironic, that the result is highly
uncertain. Indeed, we fear to fall into one of the two
possibilities: either the incident power is too low for the
nonlinear effects to be measurable, or, particularly because our
incident beam is monochromatic, the medium will not be insensitive
to light, so that thermal or mechanical effects should be included.

To conclude, the method exposed in this paper has to be generalized
in several directions: to match experimental data, we have to deal
with non-stationary media, to include non-harmonic generation (in
order to study subharmonic generation, Raman scattering, etc.), and
also to generalize our notion of the degree to the cases where the
sources of the electromagnetic field oscillate at several
frequencies (in order to study sum-harmonic generation, four-wave
mixing, etc.), or even with a continuous spectrum (in order to deal
with laser pulses).

%Nevertheless, lots of important subjects have been put aside. Among
%them, the regime where the light has an influence on the medium
%though which it propagates; in this case, the medium is not
%stationary anymore and thermal and mechanical effects are expected.
%Also, we have restrict ourself to monochromatic incident field. If
%this is possible to generalize this a bit, effects like four-wave
%mixing would be really impracticable to study with this method.
%Lastly, we did not find a notion that would generalize our notion of
%the degree when the sources of the electromagnetic field oscillate
%with a non-discrete spectrum. Unfortunately, high incident powers
%are necessary for the nonlinear effects to be significant, so that

\vspace{.5cm}%
\paragraph{Aknowledgement} The authors are grateful to S. Brasselet and A. Ferrando for their comments about this manuscript.

\bibliographystyle{unsrt}

\bibliography{biblio}

\end{document}